\documentclass[aps,prb,showpacs,twocolumn,superscriptaddress,10]{revtex4-1}
\pdfpageattr {/Group << /S /Transparency /I true /CS /DeviceRGB>>}%% to solve the color issue of Adobe reader about transparency
\usepackage{graphicx}
\usepackage{amsmath}
\usepackage{bm}

\usepackage{multirow}
\usepackage{booktabs}
\usepackage{hyperref}
\hypersetup{%
	pdftitle={Third-order optical conductivity of an electron fluid}, %
	pdfauthor={Zhiyuan Sun, D.N. Basov, and M.M. Fogler},%
	pdfpagemode={UseNone},%
	pdfstartview={FitH},%
	breaklinks=true,%
	citecolor=blue,%
	colorlinks=true,%
	linkcolor=blue,%
	urlcolor=blue}

\usepackage[T1]{fontenc}
\usepackage{fourier}
\usepackage{baskervald}

\newcommand{\unit}[1]{\,\mathrm{#1}} % for specifying units in math mode
\graphicspath{{./}}

\begin{document}

\title{Third-order optical conductivity of an electron fluid}

\author{Zhiyuan Sun}
\affiliation{Department of Physics, University of California San Diego, 9500 Gilman Drive, La Jolla, California 92093, USA}

\author{D. N. Basov}
\affiliation{Department of Physics, University of California San Diego, 9500 Gilman Drive, La Jolla, California 92093, USA}
\affiliation{Department of Physics, Columbia University,
	538 West 120th Street, New York, New York 10027}

\author{M. M. Fogler}
\affiliation{Department of Physics, University of California San Diego, 9500 Gilman Drive, La Jolla, California 92093, USA}

\date{\today}

\begin{abstract}

We derive the nonlinear optical conductivity of an isotropic electron fluid at frequencies below the interparticle collision rate.
In this regime, governed by hydrodynamics, the conductivity acquires a universal form
at any temperature, chemical potential, and spatial dimension.
We show that the nonlinear response of the fluid to a uniform field is dominated by the third-order conductivity tensor
$\sigma^{(3)}$ whose magnitude and temperature dependence
differ qualitatively from those in the conventional
kinetic regime of higher frequencies.
We obtain explicit formulas for $\sigma^{(3)}$
for Dirac materials such as graphene and Weyl semimetals.
We make predictions for
the third-harmonic generation, renormalization of the collective-mode spectrum,
and the third-order circular magnetic birefringence experiments.

\end{abstract}

\maketitle

%%%%%%%%%%%%%%%%%%%%%%%%%%%%%%%%%%%%%%%%%%%%%%%%%%%%%%%%%%%%%%%%%%%%%%%%%%%%%%%%%%%%%%%%%%%%%%%%%
%\begin{widetext}
%	\tableofcontents
%\end{widetext}

\section{Introduction}

\begin{figure}[b]
	\includegraphics[width=3.0 in]{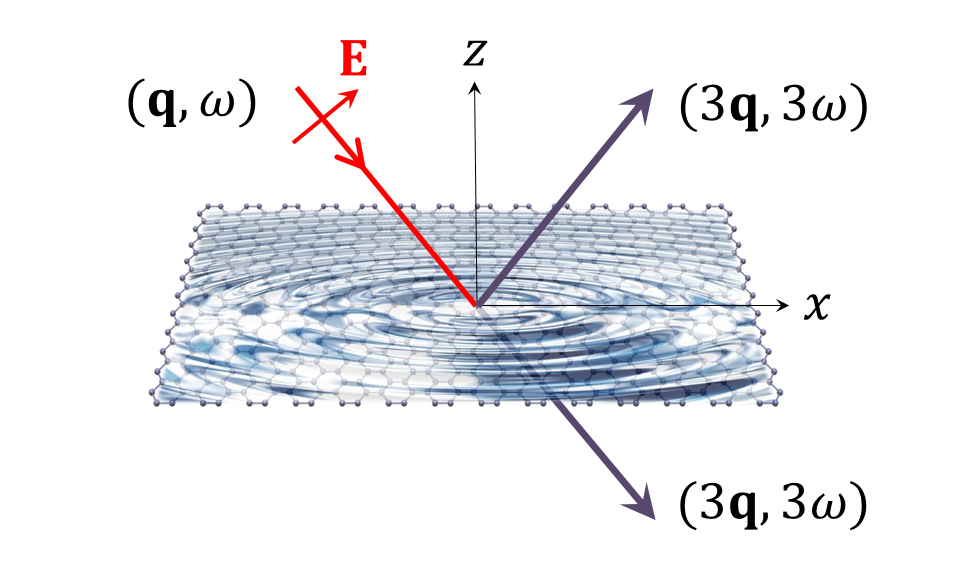}
	\caption{Illustration of the third-order optical nonlinearity in graphene.}
	\label{fig:Electron_hole_fluid}
\end{figure}

In typical metals and semiconductors electrons  experience frequent collisions with impurities and phonons. The combined rate
$\Gamma_{d}=\Gamma_{dis} + \Gamma_{ph}$ of these collisions
far exceeds the rate $\Gamma_{ee}$ of electron-electron scattering.
However, in several pure materials the opposite case $\Gamma_{d} \ll \Gamma_{ee}$ has recently shown to be possible in a range of temperatures.
Under such conditions~\cite{Gurzhi1968, Andreev2011} electrons behave as a fluid that obeys hydrodynamic equations.~\cite{Landau.6}
Evidence for the hydrodynamic behavior has been obtained from dc transport
experiments with 
two-dimensional (2D) electron gases in GaAs \cite{DeJong1995}, graphene,~\cite{Bandurin2016, Crossno2016, Kumar2017} and a quasi-2D metal $\mathrm{PdCoO_2}$.~\cite{Moll2016}
These discoveries stimulated many theoretical studies.~\cite{Muller2008a, Muller2009, Phan2013, Forcella2014, Briskot2015, Narozhny2015, Principi2015b, Principi2015, Sun2016, Lucas2016, Lucas2016b, Sun2017, Guo2017}
The conceptual simplicity of hydrodynamics
arises from dealing with only a few degrees of freedom:
the local temperature $T(\mathbf{r}, t)$, chemical potential $\mu(\mathbf{r}, t)$, and the flow velocity $\mathbf{u}(\mathbf{r}, t)$. 
The complicated many-body collisions need not be considered explicitly.
In our previous paper~\cite{Sun2017} we used this hydrodynamic formalism
to calculate the electrodynamic response of an electron fluid,
in particular,
its linear and second-order nonlinear optical conductivities.
The magnitude and functional form of these
quantities in the hydrodynamic regime of frequencies $\omega \ll \Gamma_{ee}$
were shown to differ
qualitatively from their counterparts in the conventional kinetic regime $\omega \gg \Gamma_{ee}$.
Here we continue this line of investigation by addressing the third-order nonlinearity, which controls, e.g., the third-harmonic generation (Fig.~\ref{fig:Electron_hole_fluid}), the Kerr effect, and four-wave mixing.

One reason why the third-order nonlinearity warrants attention is
dictated by the symmetry.
In general, the electrodynamic response of a conductor is characterized by the tensors $\sigma^{(n)}$
describing the 
components of the induced current proportional to the $n$\/th-power of the electric field. (The definition of
these tensors is given in Sec.~\ref{sec:Main_result}).
Unless the field is very strong, the nonlinear response of a material lacking the inversion symmetry
is dominated by the second-order conductivity $\sigma^{(2)}$.
However, in centro-symmetric systems, such as graphene,
$\sigma^{(2)}$ must vanish if the electric field is uniform, in which case
the third-order conductivity $\sigma^{(3)}$ becomes more important.

As we show below in this paper, the derivation
of the nonlinear conductivities
is straightforward within a certain model that we call the Dirac fluid.
This model is simple yet flexible enough to describe several types of solid-state materials.
The model assumes that the quasiparticles of the system
behave as Dirac fermions with the energy-momentum dispersion
$\varepsilon^2(\mathbf{p}) = m^2 v^4 + \mathbf{p}^2 v^2$.
The massless case $m = 0$ corresponds to electrons in graphene;
the massive case $m > 0$ is a reasonable approximation for
narrow-gap semiconductors.
Neglecting fermion-fermion interactions, one can readily compute the equilibrium thermodynamic parameters of 
this system,~\cite{Muller2008a, Muller2009, Briskot2015,
Sun2016, Sun2017}
such as the pressure $P = P(\mu, T)$ and the energy density $n_E = n_E(\mu, T)$.
The crucial simplification of the Dirac model is that the
energy-momentum tensor of the moving fluid
can be derived from that of the static one
by a Lorentz transformation
with $v$ in lieu of the speed of light $c$.
[In the noninteracting case,
the moving fluid is defined as the Fermi distribution
of quasiparticles with
the Doppler-shifted  energies
$\varepsilon(\mathbf{p}) - \mathbf{p} \mathbf{u}$.]
The Lorentz invariance ensures that
the hydrodynamic equations of a Dirac fluid have a simple
``relativistic''
form.~\cite{Landau.6, Muller2008a, Muller2009, Briskot2015, Narozhny2015,
Sun2016, Sun2017}
Precisely because $v \neq c$,
the solid-state systems with real Coulomb interactions are not truly Lorentz-invariant.
However,
the Dirac fluid should be a reasonable approximation
if the Coulomb interactions are not too strong,
so that $P$ and $n_E$ are dominated by the kinetic energy.
Besides graphene, examples of such Dirac fluids may include
the surface states of topological insulators
and three-dimensional Dirac/Weyl semimetals.

\begin{figure}[t]
	\includegraphics[width=3.0 in]{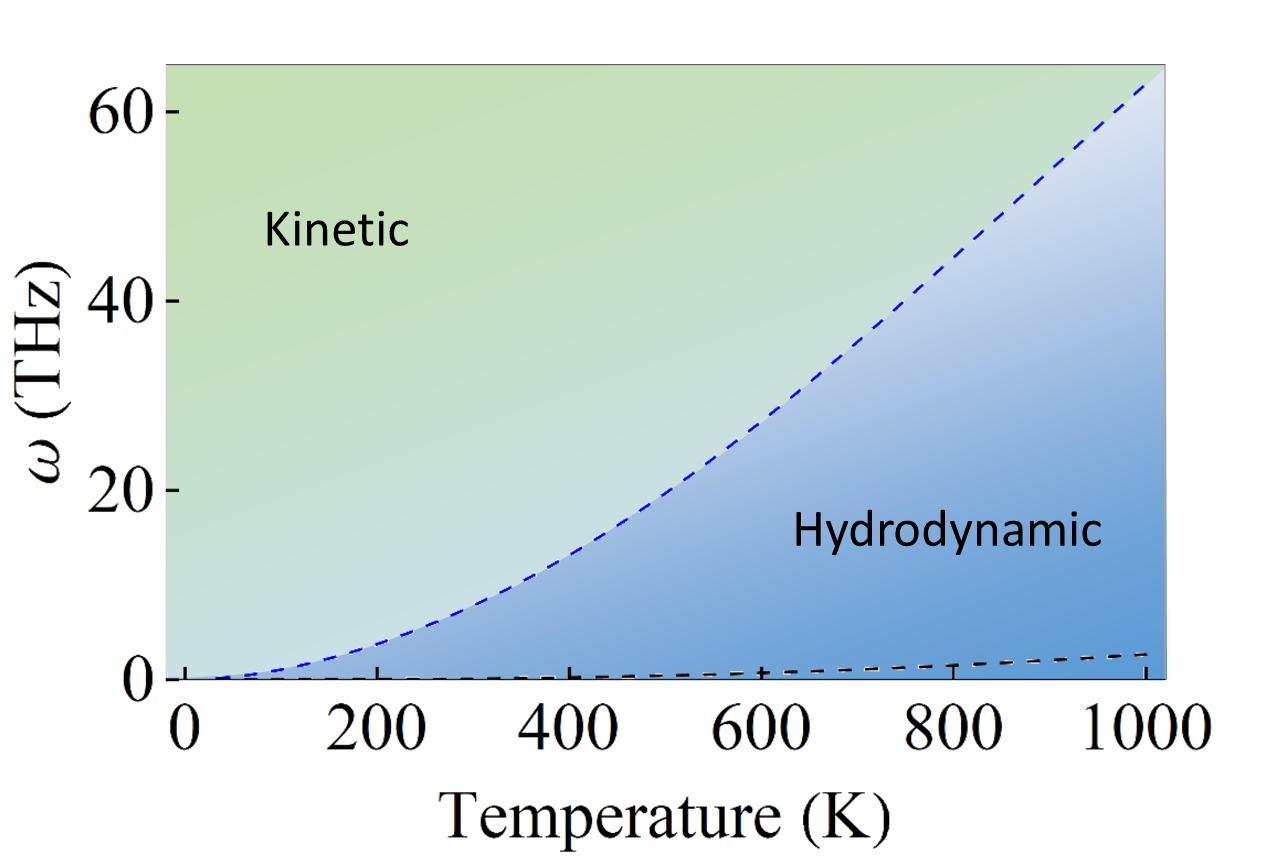}
	\caption{(Schematic) Kinetic and hydrodynamic domains in the frequency-temperature diagram of graphene for the carrier density $n=10^{12} \unit{cm^{-2}}$, corresponding to the zero-temperature chemical potential $\mu(n,0) = 0.12 \unit{eV}$.
	The (upper) dashed line	separating the two regimes is $\omega = \Gamma_{ee}(n,T)$. The lower dashed line is the momentum relaxation rate due to electron scattering by acoustic and $A_1^{\prime}$ zone-boundary phonons.
%	Note that $\mu$ decreases with $T$ for a fixed $n$.
%	The symbols represent values of $\Gamma_{ee}$ deduced from recent experiments.~\cite{Kumar2017,Bandurin2016}
}
	\label{fig:frequency_temperature}
\end{figure}

Note that the hydrodynamics regimes probed by
recent dc transport experiments~\cite{Bandurin2016, Crossno2016, Kumar2017} is less than $1$-$\mathrm{THz}$ wide.
If one wants to expand it toward higher $\omega$,
it is necessary to increase
$\Gamma_{ee}$, which can be done by raising electron temperature $T$ (Fig.~\ref{fig:frequency_temperature}).
This must be done without
heavily increasing the electron-phonon scattering rate $\Gamma_{ph}$,
which is also temperature-dependent.
One possible solution~\cite{Sun2017} is to work with (steady or transient)
states
where electrons are ``hot'' but lattice stays ``cold.''
Such nonequilibrium states can be created by optical pumping or electric-current heating.

At frequencies $\omega > \Gamma_{ee}$ where the hydrodynamic theory fails, the response of the system is better described by more conventional
approaches, e.g., the Boltzmann kinetic equation neglecting
electron-electron collisions.
Among the Dirac materials, graphene has been the most common target of such calculations. Nonlinear conductivity of graphene has been addressed
in many theoretical studies.~\cite{Mikhailov2007,
Mikhailov2011, Gullans2013, Cheng2014, Yao2014, Cheng2015, Mikhailov2016, Tokman2016, Cheng2016, Manzoni, Wang2016, Rostami2016, Sun2017}
As discussed in our previous work~\cite{Sun2017},
the differences between the hydrodynamic and kinetic regimes
(Fig.~\ref{fig:frequency_temperature}) becomes conspicuous at 
temperatures $T$ exceeding the chemical potential $\mu$,
where graphene contains two types of carriers, electrons and
thermally excited holes.
In the kinetic regime, electrons and holes tend to move in opposite directions
when driven by the electric field.
Their contributions to the electric current add up.
In the hydrodynamic regime, due to frequent interparticle collisions, all
the carriers tend to move together.
Hence, the electron and hole currents partially cancel.
As a consequence, there is an increased effective mass per unit charge, resulting in reduced linear and second-order nonlinear conductivities.\cite{Sun2017}
In this work we show that the third-order
electrodynamic response also exhibits distinct behaviors in the two regimes,
in accord with this physical picture.

The remainder of the paper is organized is follows.
Section \ref{sec:Main_result} gives a summary of our main results
such as the analytical formula for the third-order nonlinear conductivity of an isotropic Dirac fluid.
This formula is simple and universal. It is valid for any mass $m$, chemical potential $\mu$, temperature $T$, and space dimension $d$
if momentum nonconserving processes can be neglected, $\Gamma_d \to 0$.
We also discuss the general form of the higher nonlinear conductivities tensors $\sigma^{(n)}$ of odd order $n$.
In Sec.~\ref{sec:hydro} we introduce the relativistic hydrodynamic equations and apply them to the massless case, such as graphene.
In Sec.~\ref{sec:sigma3} we give the derivation of $\sigma^{(3)}$,
including the case of an external applied magnetic field.
In Sec.~\ref{sec:THG} we apply our results for $\sigma^{(3)}$
to computing the third harmonic generation.
In Sec.~\ref{sec:demon} we discuss the Kerr effect and
its influence on the hydrodynamic collective modes of the fluid.
In Sec.~\ref{sec:cd} we compute the magnetic-field-induced third-order circular birefringence.
The concluding remarks are given in Sec.~\ref{sec:discussion}.
Appendix~\ref{appendix:thermodynamics} provides a summary of the analytical expressions for the thermodynamic quantities of a massless Dirac fluid.
Appendix~\ref{appendix:derivation} outlines the derivation of $\sigma^{(3)}$ for a more realistic case of a finite scattering rate $\Gamma_d$.

\section{Main results}
\label{sec:Main_result}

The $N$\/th order nonlinear ac conductivity $\sigma^{(N)}$ is defined as a rank $(1,N)$ tensor which maps electric fields to the $N$\/th order electrical current
\begin{align}
j_{i}^{(N)}(\mathbf{q}, \omega)
&=\sum\limits_{\nu_1 ... \nu_{N}} \, \sum\limits_{(\mathbf{q}_1,\omega_1),(\mathbf{q}_2,\omega_2)\,\ldots (\mathbf{q}_N,\omega_N)} 
\delta \left( \sum_{a}^{N} (\mathbf{q}_a,\omega_a) - (\mathbf{q}, \omega) \right) \notag\\
&\sigma^{(N)}_{i \nu_1 ... \nu_{N}} (\mathbf{q}_1,\omega_1,\mathbf{q}_2,\omega_2\,\ldots, \mathbf{q}_N,\omega_N) \notag\\
& E_{\nu_1}\left(\mathbf{q}_1, \omega_1\right) E_{\nu_2}\left(\mathbf{q}_2, \omega_2\right) ...
E_{\nu_{N}}\left(\mathbf{q}_N, \omega_N\right)\,.
\label{eqn:nonlinear_conductivity}
\end{align}
The even, e.g., second order conductivities vanish at zero momentum in centro-symmetric systems. We therefore focus on odd order (e.g., $N = 3$) conductivities and disregard $\mathcal{O}(q^2)$ nonlocal corrections.

For a general $d$-dimensional charged ideal fluid with $O(d)$ (rotation and reflection) symmetry, the $N$\/th order nonlinear optical conductivity is found to be
\begin{align}
	\sigma^{(N)}_{i \, \nu_1 \nu_2 \,\ldots,\nu_{N}} = \frac{i D_h^{(N)}}{\omega_1 \omega_2 \,\ldots, \omega_{N}} \Delta_{i \, \nu_1 \nu_2 ... \nu_{N}}
	\,
	\label{eqn:sigma_odd}
\end{align}
where 
\begin{align}
\Delta_{i\,\nu_1 \nu_2 ... \nu_{N}} = \delta_{i\,\nu_1}\delta_{\nu_2 \nu_3}...\delta_{\nu_{N- 1} \nu_{N}} + \mathrm{permutations}
\,
\label{eqn:delta_tensor}
\end{align}
is the totally symmetric rank $N + 1$ tensor which is the sum of the $N!!$ isotropic tensors. Note that $O(d)$ symmetry only requires that $\sigma^{(N)}$ is a linear combination of the isotropic tensors. As we will show later, due to the additional condition of thermal equilibrium in the hydrodynamic regime, $\sigma^{(N)}$ can only be proportional to the totally symmetric rank $N + 1$ tensor $\Delta_{i, \nu_1 \nu_2 \,\ldots, \nu_{N}}$. The hydrodynamic $N$\/th order optical weight $D_h^{(N)}$ should be understood as a thermodynamic quantity which is generally unknown.

Applied to the Dirac fluid, which has (quasi) Lorentz symmetry, the linear optical conductivity is recovered as $\sigma_{ij}=\frac{iD_h/\pi}{\omega} \delta_{ij}$ where the hydrodynamic Drude weight is $D_h=\pi ne^2/m^{\ast}$ (see, e.g., Supplemental material of Ref.~\onlinecite{Sun2017}).
The most important result of this paper is that the third-order nonlinear optical conductivity is given by
\begin{align}
\sigma^{(3)}_{ilmn} &=\frac{i D_h^{(3)}}{\omega_1 \omega_2 \omega_3} \Delta_{ilmn} \notag \\
& = \frac{i D_h^{(3)}}{\omega_1 \omega_2 \omega_3}
\left(\delta_{il}\delta_{mn} + \delta_{im}\delta_{ln} + \delta_{in}\delta_{lm}\right)
\,.
\label{eqn:sigma_3}
\end{align}
Here the third-order optical weight
\begin{align}
D_h^{(3)} = \frac{1 - C_{\mathrm{ise}}}{3!} \frac{e^4 n}{m^{\ast 3} v^2} =\frac{1 - C_{\mathrm{ise}}}{3!} \frac{W}{\rho^4 v^4} \left(\frac{D_h}{\pi}\right)^4
\,.
\label{eqn:D_h}
\end{align}
is expressed in terms of thermodynamic quantities and the asymptotic velocity $v$.
These quantities are defined in Sec.~\ref{sec:hydro} below.

\section{Hydrodynamics}
\label{sec:hydro}

\subsection{Hydrodynamic equations}

The hydrodynamic equations for an ideal relativistic charged fluid are\cite{Landau.6}
\begin{align}
	\partial_\mu T^{\mu\nu} =  J_\mu F^{\nu\mu} ,\quad \partial_\mu J^\mu=0 \,
	\label{Eq_hydro1}
\end{align}
where $T^{\nu\mu}$ is the energy-momentum tensor, $F^{\nu\mu}$ is the electromagnetic field tensor,
and
\begin{align}
J^\mu=(\rho v,\, \mathbf{j}\,),
\qquad \mathbf{j} = (j_x, j_y, j_z)
	\label{Eq_Jj}
\end{align}
is the four-current and its spatial part, respectively.
By ideal we mean a fluid with vanishing viscosity and thermal conductivity.
Although a ``covariant'' notation is implemented, Eq.~\eqref{Eq_hydro1} holds even for systems without Lorentz symmetry because the conservation of the stress tensor
requires only the translational symmetry. The form of stress tensor
for a general interacting fluid is unknown; however, 
for Lorentz-invariant systems, i.e., Dirac fluids, the stress tensor is related to thermodynamic quantities\cite{Landau.6}
\begin{align}
	T^{\mu\nu} = Wu^\mu u^\nu - P g^{\mu\nu} \,.
	\label{Eq_stress_tensor}
\end{align}
The four-current is related to the proper charge density through $J^{\mu}=\rho_0 u^{\mu} = \rho_0 \gamma (v,\, \mathbf{u})=\rho ( v,\, \mathbf{u})$ where $\gamma=1/\sqrt{1-u^2/v^2}$. In solid state system, the $v$ is the asymptotic velocity such that electrons have a Dirac-like energy-momentum dispersion
$\varepsilon_p^2 = (p v)^2 + (m v^2)^2$. The thermodynamic quantities, the proper charge density $\rho_0= e n_0$, the enthalpy density $W$ and the pressure $P$ are all defined in the fictitious proper frame moving with the local liquid. Note that $n \equiv \rho/e$ is defined as the effective charge carrier density and is in general not the same as particle density. For example, in graphene at high temperature, there are both electrons and holes, and $n$ will be the number of electrons minus the number of holes. We define the hydrodynamic effective mass as $m^{\ast} \equiv W/(nv^2)$. 
And we define 
\begin{align}
	C_{\mathrm{ise}}= \frac{n}{W} \left(\frac{\partial W}{\partial n}\right)_{\mathrm{ise}} -1 = \frac{n}{W}
	\left( \frac{\partial P}{\partial n} \right)_{\mathrm{ise}}
	= \frac{1}{m^{\ast}\! v^2}
	\left( \frac{\partial P}{\partial n} \right)_{\mathrm{ise}}
\end{align}
as the dimensionless bulk isentropic modulus of the electron fluid\cite{Sun2017}. For example, it has the value $1/d$ for massless Dirac particles in space dimension $d$. 
Out of the three thermodynamic quantities $W$, $P$ and $\rho_0$ only two are independent.
Thus the independent variables are any two thermodynamic quantities and the local flow velocity $\mathbf{u}$. This set of equations \eqref{Eq_hydro1} is closed.

Alternatively, these hydrodynamic equations can be derived from the Boltzmann kinetic equation with the inter particle collision integral but neglecting the many-body interaction correction to the thermodynamic quantities, as shown in Ref.~\onlinecite{Briskot2015} and also the covariant version in Appendix \ref{appendix:RBE}.

From Eq.~\eqref{Eq_stress_tensor}, the first part of Eq.~\eqref{Eq_hydro1} could be written in another form
\begin{align}
Wu^{\mu} \partial_{\mu} u^{\nu} -\partial^{\nu}P + u^{\nu} u^{\mu} \partial_{\mu}P  =
J_{\mu} F^{\nu\mu} \,.
\label{Eq_hydro2}
\end{align}
Separating the time and spatial components in a proper way, Eq.~\eqref{Eq_hydro1} has another form
\begin{align}
	  & (\partial_t + u_k\partial_k) u_i =  \, \notag                                                                                                                                    \\
	  & \frac{1}{\gamma^2 W} \left( -\partial_i P - u_i \partial_t P + \rho E_i + \rho\epsilon_{ikl}u_k \frac{v}{c} B_l - u_i\mathbf{j} \cdot \mathbf{E}  \right)   \,, \label{Eq_euler} \\
	  & \partial_t (n_{E}) + \nabla(\gamma^2 W \mathbf{u})  =  \mathbf{j} \cdot \mathbf{E}  \,,
	\label{Eq_energy_c}\\
	  & \partial_t \rho + \nabla \cdot \mathbf{j}  =0 \,.
	\label{Eq_charge_c}
\end{align}
where $n_{E}=\gamma^2 W - P$ is the energy density of the electrons (relative to zero doping and temperature case), and $n_{E0}=W-P$ is the same quantity but in the proper frame. Equation~\eqref{Eq_euler} is the relativistic version of the Euler equation, Eq.~\eqref{Eq_energy_c} is the conservation of the energy current and Eq.~\eqref{Eq_charge_c} is the conservation of the charge current. Terms due to viscosity and dissipative thermal conductivity are neglected because they affect the conductivities only through $O(q^2)$ terms.
To simplify the notations, the asymptotic velocity has been taken to be $v=1$ except for the Lorentz force term due to the magnetic field $\mathbf{B}$.
Note that the magnetic field is related to the electric one by $\nabla \times \mathbf{E} = - c^{-1} \partial_t \mathbf{B}$. Since we neglect finite-$q$ effects in this work, we must set
$\nabla \times \mathbf{E} = 0$, so that the magnetic field is considered time-independent.

Starting from Eqs.~\eqref{Eq_euler}, \eqref{Eq_energy_c} and \eqref{Eq_charge_c} we
can compute the linear and in principle, any higher-order nonlinear optical conductivities $\sigma^{(N)}$
by expanding all the dynamic variables in powers of the electric field $\mathbf{E}$.
This procedure and its results are presented in the following sections.

\subsection{Hydrodynamic regime of graphene}
\label{Sec:hydro_regime_graphene}

Hydrodynamic regime is not a phase of matter but a domain of
frequency-momentum diagram [see, e.g., Fig.~1 of Ref.~\onlinecite{Sun2017}]
where the hydrodynamic equations work well as an effective theory. This regime is defined by inequalities $\Gamma_d,\, \omega\ll\Gamma_{ee}$ and $q \ll l_{ee}^{-1}$,
which can be satisfied in some pure solid-state systems.
The electron-electron collision rate $\Gamma_{ee}(n,T)$ that sets the upper bound on the hydrodynamic regime depends on temperature and doping level of the electron system.
For example, in graphene,
$\Gamma_{ee}$ scales as $\sim \ln (2\mu/T) (T^2/\mu)$ at low $T$ and as $\sim \alpha^2 T$ at $T\gg\mu$ with $\alpha \sim 1$. 
For a rough estimate, we connect these two formulas by a
naive interpolation with the relative weights $\mu / (T + \mu)$ and $T / (T + \mu$),
respectively.
The corresponding boundary of the hydrodynamic domain (blue region) is shown by the upper dashed line in Fig.~\ref{fig:frequency_temperature}.
The other important scattering rate shown in the same Figure is $\Gamma_{d}$. For ultra clean graphene encapsulated in hexagonal boron nitride (hBN), the major contribution to $\Gamma_{d}$ is the electron-phonon scattering  $\Gamma_{ph}$. The electron-phonon scattering suppresses the hydrodynamic behavior if $\Gamma_{ph} > \Gamma_{ee}$. However, our theoretical estimation of $\Gamma_{ph}$ and $\Gamma_{ee}$ indicates that the hydrodynamic regime of graphene is fairly large, as shown in Fig.~\ref{fig:frequency_temperature}. Furthermore, in a nonequlibrum situation where the electron system is at high temperature $T$ while the lattice temperature $T_l$ remains low, $\Gamma_{ee}$ is enhanced while the $\Gamma_{ph} = \Gamma_{ph}(T_l,T,n)$ remains small, and so the hydrodynamic window could be wider. Such a nonequlibrum situation can be realized either through optical pumping~\cite{Ni2016} or by Joule heating due to an electric current.~\cite{Son2017}

Recent experiments~\cite{Bandurin2016, Crossno2016, Kumar2017,Moll2016} explored the dc transport in several pure 2D conductors in the regime $\Gamma_d\ll\Gamma_{ee}$, i.e., near the horizontal axis $\omega = 0$ of Fig.~\ref{fig:frequency_temperature}. These measurements revealed
signatures of viscous electron flow, which manifest themselves through a combination of high viscosity $\nu = v^2 / \Gamma_{ee}$ and geometrical restriction on the flow. Because of the particular focus of these studies, the lower viscosity, higher temperature regime was labeled as nonhydrodynamic. From our point of view, viscosity is just one of very many hydrodynamic phenomena rather than its essential element. Actually, in the high-temperature regime, where the electron-hole plasma becomes a more perfect fluid,\cite{Muller2009} hydrodynamic effects should be of crucial importance. They are predicted to give dramatically different optical responses compared to the noninteracting kinetic theory.~\cite{Sun2017}

\section{Third-order nonlinear optical conductivity}
\label{sec:sigma3}

\subsection{The general charged fluid}
Since we neglect the $\mathcal{O}(q^2)$ nonlocal corrections, $\sigma^{(N)}$ could be derived by simply considering the fluid driven by a uniform electric field. (The dynamic magnetic field $\mathbf{B}$ is zero in this case.)
The hydrodynamic equations~\eqref{Eq_hydro1} simplify to
\begin{align}
	\partial_t \mathbf{p} = \rho \mathbf{E}\,,
	\quad
	\partial_t \rho =0 \,,\quad \partial_t S_n=0  \,.
	\label{Eq_hydro_uniform}
\end{align}
where $\mathbf{p}$ is the momentum density and
$S_n = S/n$ is the entropy per unit charge. The last relation in Eq.~\eqref{Eq_hydro_uniform} comes from the fact that the hydrodynamic flow is isentropic. 
The second relation in Eq.~\eqref{Eq_hydro_uniform} comes
from the charge continuity equation. It entails that the charge density $\rho$ stays constant.
In turn, the first relation in Eq.~\eqref{Eq_hydro_uniform}
implies that $\mathbf{p}$ is strictly linear in $\mathbf{E}$.
If the electric field in the system is composed of
Fourier harmonic with amplitudes $\mathbf{E}_a$ and frequencies $\omega_a$,
then the momentum density is
\begin{align}
\mathbf{p} = \sum_{a = 1}^{N} \frac{i}{\omega_a}
\rho \mathbf{E}_a e^{-i\omega_a t} \,.
\label{Eq_p_E}
\end{align}
The current density can be treated as a thermodynamic function of $n$, $S_n$ and $\mathbf{p}$:
\begin{align}
\mathbf{j} = \mathbf{j}(n,S_n,\mathbf{p})  \,.
\label{Eq_momentum}
\end{align}
Since particle density and entropy are conserved, the $N$\/th order current 
where $N = 2m + 1$, is just the $N$\/th order Taylor expansion of $\mathbf{j}$ with respect to $\mathbf{p}$. Due to the isotropy of the fluid, current density must be parallel to the momentum:
$\mathbf{j}= j \hat{\mathbf{p}}$.
Therefore,
\begin{align}
j_i^{(N)} = \frac{1}{N!}\left( \partial_p^{N} j \right) (p^2)^m p_i  \,.
\label{Eq_j_p}
\end{align}
Using Eq.~\eqref{Eq_p_E},
we find the $N$\/th order current of frequency $\omega=\sum_{a}^{N}\omega_a$ to be
\begin{align}
j_{i}^{(N)} &= \frac{1}{N!}\left( \partial_p^{N} j \right)  \frac{i(-1)^m}{\omega_1 \omega_2 \ldots \omega_{N}} \delta_{i\,\nu_1 }\delta_{\nu_2 \nu_3}...\delta_{\nu_{N-1} \nu_{N}} \notag\\
& \times E_{1,\nu_1} E_{2,\nu_2} \ldots
 E_{N,\nu_{N}}  + \mathop{\mathrm{perm}} (1,2,\ldots,N) \,.
\label{Eq_j_p}
\end{align}
Here and below ``$\mathrm{perm}$'' stands for permutations.
Therefore, Eq.~\eqref{eqn:sigma_odd} is proven with 
\begin{align}
D_h^{(N)}=  \frac{(-2)^m m!}{(N!)^2}  \left(\partial_p^{N} j \right)\,.
\label{Eq_D_h_n}
\end{align}

\subsection{The Lorentz invariant fluid (Dirac fluid)}
As we mentioned above, Eq.~\eqref{Eq_hydro_uniform} implies that the charge density $\rho$ is a constant. From Lorentz invariance, the current is $j_i = \rho u_i$ and therefore $j^{(3)}_i = \rho u^{(3)}_i$. The flow velocity $\mathbf{u}$ can be found from its nonlinear relation to the momentum $p_i = W \gamma^2 u_i$. The left hand side is linear in electric field, thus the third-order terms on the right hand side must vanish:
\begin{align}
	0 = Wu_i^{(3)} + W^{(2)}u_i^{(1)} + W {u^{(1)}}^2 u_i^{(1)} \,.
	\label{Eq_third_order_simple}
\end{align}
It follows that
\begin{align}
	u_i^{(3)} & = - \frac{ W^{(2)}}{W}u_i^{(1)} - {u^{(1)}}^2 u_i^{(1)} =  -\left( \frac{(\frac{\partial W}{\partial n_0})_{S_n} n^{(2)}_0}{W} + {u^{(1)}}^2 \right) u_i^{(1)} 
	\label{Eq_u_3_1} \,.
\end{align}
From the relativistic relation $n=\gamma n_0$, we have
\begin{align}
 n_0=n \left( 1-{u}^2/2 + O({u}^4) \right) \,,
\end{align}
Thus $n_0^{(2)}= n {u^{(1)}}^2/2$ and
\begin{align}
u_i^{(3)}  & = \frac{1}{2} \left(C_{\mathrm{ise}} -1 \right) {u^{(1)}}^2 u_i^{(1)} \,.
\label{Eq_third_order}
\end{align}
Therefore, we arrive at
\begin{align}
j^{(3)} = \rho u^{(3)} = \frac{1}{2} \left(C_{\mathrm{ise}} -1 \right) (p/W)^3
\label{Eq_third_order_j}
\end{align}
which renders the third-order optical weight Eq.~\eqref{eqn:D_h}.

\subsection{2D Dirac fluid with a static magnetic field}
\label{sec:magnetic}

The uniform hydrodynamic equations with a static magnetic field are
\begin{align}
\partial_t p_i = \rho E_i + \rho \frac{1}{c} \epsilon_{ijk} u_j B_k  \,,\quad \partial_t \rho =0 \,,\quad \partial_t S_n=0  \,.
\label{Eq_momentum}
\end{align}
The momentum density $\mathbf{p}$ is no longer strictly linear in $\mathbf{E}$. Below we focus on the 2D case, so the Dirac fluid is on $x$-$
y$ plane and the static magnetic field is in the $z$-direction. The linear conductivity is modified to~\cite{Muller2008a}
\begin{align}
%\sigma_{ij} &= \frac{1}{-(\frac{\omega}{D_h/\pi})^2 + \frac{B^2}{\rho^2 c^2}} \left( \frac{-i\omega}{D_h/\pi} \delta_{ij} + \frac{B}{\rho c} \epsilon_{ij} \right) \,\notag\\
\sigma_{ij} = \frac{D_h/\pi}{\omega^2 -\omega_c^2} \left( i \omega \delta_{ij} - \omega_c \epsilon_{ij} \right)
\label{Eq_sigma}
\end{align}
where $\epsilon_{ij}$ is the antisymmetric tensor in 2D, and $\omega_c = eB/(m^{\ast} c)$ is the hydrodynamic cyclotron frequency. Note that $\omega_c$ is in general not equal to the usual cyclotron frequency because the hydrodynamic effective mass $m^{\ast} \equiv W/(nv^2)$ is not exactly the same as the quasiparticle effective mass in a Fermi liquid.  From the Euler equation, the third-order momentum is related to the third-order flow velocity
\begin{align}
-i\omega_s p_i^{(3)} =  \rho \frac{1}{c} \epsilon_{ijk} u_j^{(3)} B_k   \,.
\label{Eq_momentum}
\end{align}
where $\omega_s$ is the frequency of the third-order current (the sum frequency). Together with the relation 
\begin{align}
p_i^{(3)}= Wu_i^{(3)} + \frac{1}{2}(1-C_{\mathrm{ise}}) W (u^{(1)})^2 u_i^{(1)}
\end{align}
we get the equation for the third-order current
\begin{align}
M_{ij} j_j^{(3)}= \frac{1}{2}(1-C_{\mathrm{ise}}) \frac{W}{\rho^3} (j^{(1)})^2 j_i^{(1)}
\end{align}
where $\hat{M}= -\frac{i}{\omega_s} \rho \hat{\sigma}^{-1}$, therefore 
\begin{align}
j_i^{(3)} &= \frac{1}{2}(1-C_{\mathrm{ise}}) \frac{W}{\rho^4} i\omega_s \sigma_{il}(\omega_s) \left(j^{(1)} \right)^2 j_l^{(1)} \notag \\
&=\frac{1}{2}(1-C_{\mathrm{ise}}) \frac{W}{\rho^4} i\omega_s \sigma_{i\alpha}(\omega_s) \sigma_{\alpha l}(\omega_1) \sigma_{k m}(\omega_2) \sigma_{k n}(\omega_3) \notag\\
& \times E_{1 l} E_{2 m} E_{3 n} + \mathop{\text{perm}} (1,2,3)\,.
\end{align}
The symmetrized third-order conductivity reads
\begin{align}
\sigma^{(3)}_{ilmn} &= \frac{1}{3! \cdot 2} (1-C_{\mathrm{ise}}) \frac{W}{\rho^4} i\omega_s \sigma_{i\alpha}(\omega_s) \sigma_{\alpha l}(\omega_1) \sigma_{k m}(\omega_2) \sigma_{k n}(\omega_3) \notag\\
&+ \mathop{\text{perm}} (1,2,3)(l,m,n)
\label{eqn:sigma3_sigma1}
\end{align}
where $\mathop{\text{perm}} (1,2,3) (l,m,n)$ denotes the $3! = 6$ permutations of the indices $(l,m,n)$ together with $(1,2,3)$. Therefore, we have proven that the third-order conductivity $\sigma^{(3)}$ is determined by the linear one $\sigma$.

For moderate magnetic field $\omega_c \ll \omega$, and the case of a single frequency $\omega_1=\omega_2=\omega_3=\omega$, we can expand $\sigma^{(3)}$ to linear order in $B$: 
\begin{align}
\sigma^{(3)}_{ilmn}= \frac{i D_h^{(3)}}{\omega^3} \left[ \Delta_{ilmn} +i \frac{4 \omega_c}{3 \omega} \Xi_{ilmn} \right] ,
\label{eq:sigma3_B}
\end{align}
where $\Xi_{ilmn}=\delta_{lm}\epsilon_{in}+\delta_{ln}\epsilon_{im}+\delta_{mn}\epsilon_{il}$.

\subsection{Analysis of $\sigma^{(3)}$}
The simple tensorial structure of Eq.~\eqref{eqn:sigma_3} is a result of rotational symmetry and local equilibrium nature of an ideal charged fluid. For comparison, in the high frequency kinetic/quantum regime of graphene, the tensorial structure of $\sigma^{(3)}$ is more complicated due to contributions from interband transitions and disorder scattering effects.~\cite{Cheng2014,Cheng2015,Mikhailov2016}
In the hydrodynamic regime, the interband transitions are suppressed by fast e-e scattering, resulting in the simple expression of Eq.~\eqref{eqn:sigma_3}.

The magnitude of hydrodynamic $\sigma^{(3)}$ is also different from that of the kinetic theory. Applied to graphene at $T = 0$, our result
for $D_h^{(3)}$ is
\begin{align}
	D_h^{(3)}(T=0) = \frac{g}{48\pi} \frac{e^4 v_F}{\hbar^3 k_F} = 2 D_k^{(3)}
	\,,
	\label{eqn:D_h_zero_t}
\end{align}
which is twice the third-order spectral weight $D_k^{(3)}$ from the collisionless Boltzmann transport theory \cite{Mikhailov2016,Cheng2014,Mikhailov2007}. This difference could be measured by, e.g., third harmonic generation to be discussed in Sec.~\ref{sec:THG}.

Comparing the third-order nonlinear response with the usual linear one,
we notice that the third-order current is suppressed by the parameter
\begin{align}
\xi = \left( \frac{-eE/\omega}{m^{\ast} v} \right)^2 \ll 1\,.
	\label{eqn:xi}
\end{align}
This factor is different in the nonrelativistic and the ultrarelativistic regimes because $m^*$ depends
on the Fermi momentum $p_F$.
As a result, in the nonrelativisitic case parameter $\xi$ is smaller by the factor of $(v_F / v)^2 \ll 1$
than the ultrarelativistic case.
This factor vanishes for a system with the parabolic dispersion, which
corresponds to $v \to \infty$.
Indeed, for such a system all nonlinearities at zero $q$ should be absent because of
the Galilean invariance (Kohn's theorem).
On the other hand, in graphene at zero temperature, which is an example of the ultrarelativistic system,
$m^{\ast} v = p_F$, so that $\xi = ({\delta p} / {p_F})^2$.
Here $\delta p = -e E /\omega$ has the physical meaning of the amplitude of electron momentum oscillation caused by the electric field.

The third-order conductivity \eqref{eqn:sigma_3} diverges at the zero frequency limit, which is unphysical. In reality,
the divergence is curbed by the momentum relaxation rate $\Gamma_d$, similar to the first order conductivity. A simple
but crude way to include the effect of the momentum relaxation is to change all the frequencies $\omega_a$ to $\omega_a^+ = \omega_a + i\Gamma_d$. However, this approach neglects the increase of entropy density due to momentum relaxation.
Thus, special care needs to be taken to compute the true nonlinear dc response, as shown in Appendix \ref{appendix:derivation}.

\section{Third harmonic generation}
\label{sec:THG}
\begin{figure}
	\includegraphics[width=3.0 in]{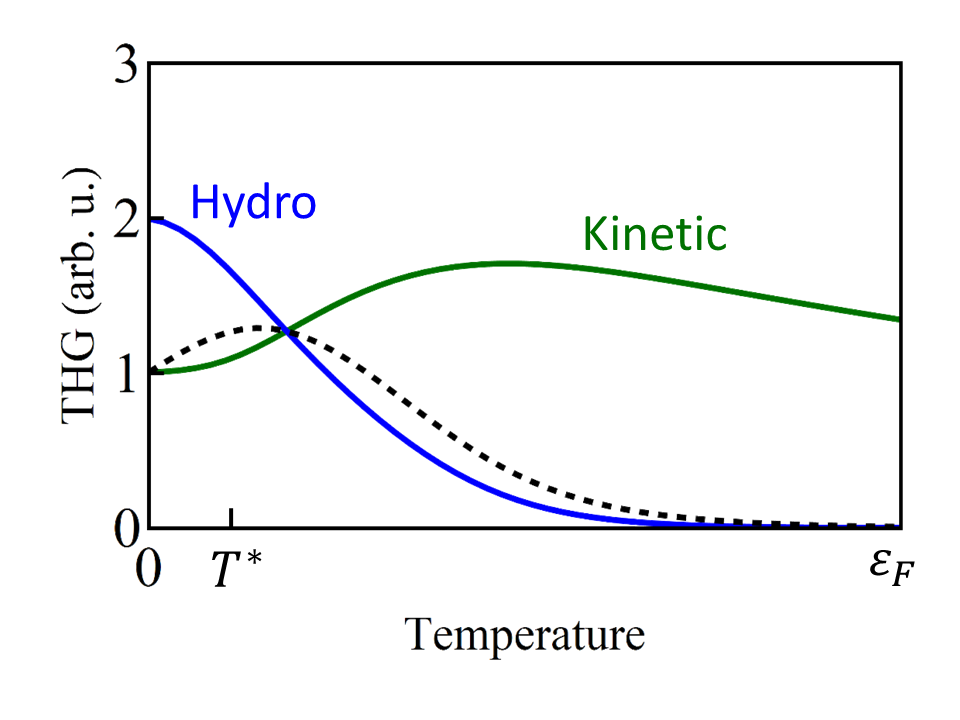}
	\caption{The THG signal as a function of temperature at fixed $n$. The frequency of incident light is $\hbar \omega = \varepsilon_F/20$ with $\varepsilon_F$ being the zero temperature Fermi energy. The blue curve is the prediction of the hydrodynamic theory Eq.~\eqref{eqn:sigma_3}, the green curve is from the RPA.~\cite{Cheng2015} The observable signal should behave as sketched by the dashed curve.}
	\label{fig:THG}
\end{figure}
One quantity we can derive from $\sigma^{(3)}$ [Eq.~\eqref{eqn:sigma_3}] is the third harmonic generation (THG) \cite{Mikhailov2016,Lupi.2016}.
Assume the ac electric field of the incident light is in the $x$-direction 
\begin{equation}
\mathbf{E}(t) = \hat{x}E(\omega) e^{-i \omega t} + \mathrm{c.c.} 
\label{eqn:E}
\end{equation}
The current, which determines the observable THG signal is
\begin{equation}
\mathbf{j}^{(3)}(t) = \hat{x} \sigma^{(3)}_{xxxx}(\omega,\omega,\omega) E(\omega)^3 e^{-i 3\omega t} + \mathrm{c.c.} 
\label{eqn:E}
\end{equation}
Therefore, $\sigma^{(3)}_{xxxx}(\omega,\omega,\omega)$ represents the magnitude of the THG. 
This quantity is plotted in Fig.~\ref{fig:THG} as a function of $T$.
Also shown in Fig.~\ref{fig:THG} is the prediction of the conventional theory based on random-phase approximation (RPA),~\cite{Cheng2015, Mikhailov2016} which is valid in the kinetic regime.
The two curves exhibit different behavior. At zero temperature, the hydrodynamic theory predicts the THG signal which is twice that of the kinetic theory. However, since $\Gamma_{ee}=0$ at $T=0$, the electron system much be in the kinetic regime (Fig.~\ref{fig:frequency_temperature}), and so the actual $\sigma^{(3)}_{xxxx}(\omega,\omega,\omega)$ should be close to the kinetic theory value, as sketched by the dashed line. As temperature increases, $\Gamma_{ee}=0$ grows, and the system will experience a crossover from the kinetic to hydrodynamic regime at certain $T^{\ast}$. This crossover temperature is determined by $\Gamma_{ee}(n,T^{\ast})=\omega$. As temperature increases further, the hydrodynamic third-order optical weight drops as $D^{(3)}_{h} \propto (m^{\ast})^{-3} \propto T^{-9}$
due to the thermal enhancement of the hydrodynamic effective mass $m^{\ast}(n,T)$.
Therefore, the THG drops much faster than what the conventional kinetic theory would predict.  

\section{The Kerr effect and the demons}
\label{sec:demon}

\begin{figure}[b]
	\includegraphics[width=2.6 in]{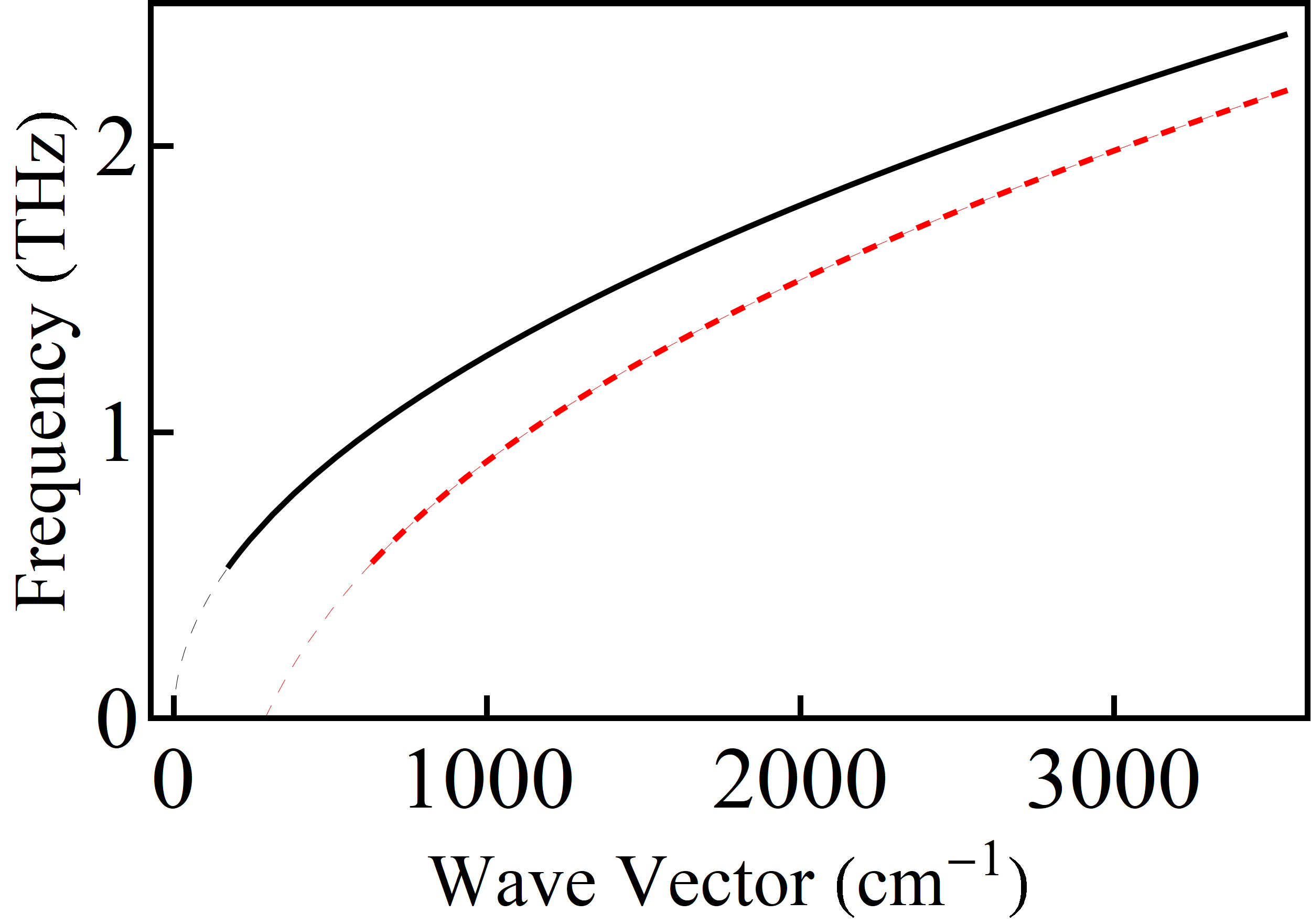}
	\caption{The dispersion of the demons in the hydrodynamic regime of graphene. The black curve is for the weak field limit while red dashed curve includes the Kerr-effect-induced shift in a strong field ($E=10^4 \unit{V/cm}$). The carrier density and temperature are $n = 10^{12} \unit{cm^{-2}}$ and $T = 300 \unit{K}$.}
	\label{fig:Kerr}
\end{figure}

The Kerr effect refers to the change of the effective permittivity of a medium due to the third-order nonlinearity\cite{Landau.8,Boyd.2008}. For a 2D charged Dirac fluid, this effect is more conveniently described as the shift of the effective conductivity. One manifestation of the Kerr effect is the renormalization of the frequency of the collective modes in a strong optical field. In the kinetic regimes these modes
are the familiar plasmons.~\cite{Mikhailov2017,Lupi.2016}
In the hydrodynamic regime, they are the demons.~\cite{Sun2016,Sun2017}

In general, to describe the collective modes, we need to study response at a finite $q$. For small enough $q$, we can approximate
the result using $q = 0$ quantities, as follows.
The charge density fluctuation could be represented by means of the Fourier amplitude $\rho_{\omega,q}$:
\begin{align}
	\rho &= \rho_{\omega,q} e^{i(\mathbf{q}\cdot \mathbf{r} -\omega t)} + \text{c.c.}
\end{align}
Assume $q$ is in the $\hat{x}$ direction. The corresponding Fourier amplitude of the electric field is $E_x=(-iq)v_q \rho_{\omega,q}$, where $v_q$ is the Coulomb potential. In 2D, it is given by $v_q = 2\pi / \kappa q$. The electric field induces the current
\begin{align}
	j_x(\omega) &= \sigma(\omega)E_x + 3\sigma^{(3)}_{xxxx}(\omega,\omega,-\omega) E_x E_x E_x^{\ast} 
	\,.
\end{align}
Using the charge continuity equation $\partial_t \rho + \nabla \mathbf{j} = 0$, we obtain
\begin{align}
	-i\omega \rho + q^2 v_q \sigma(\omega) \rho + 3 q^4 v_q^3 \sigma^{(3)}_{xxxx}(\omega,\omega,-\omega) \rho \rho \rho^{\ast} =0 
	\,.
\end{align}
The weak-field dispersion can be obtained from this equation by dropping the last term.
When this term is retained, the dispersion acquires the frequency shift
proportional to $\sigma^{(3)}$:
\begin{align}
	\delta\omega &= -\frac{3i}{2} q^4 v_q^3 \sigma^{(3)}_{xxxx}(\omega,\omega,-\omega) |\rho_{\omega,q}|^2 \notag\\
	&=-\frac{3i}{2} q^2 v_q \sigma^{(3)}_{xxxx}(\omega,\omega,-\omega) |E_{\omega,q}|^2 
	\,.
\end{align}
[To obtain this relation we also assumed that the linear conductivity has the Drude form $\sigma(\omega) \propto \omega^{-1}$.]
Applied to Eq.~\eqref{eqn:sigma_3}, we obtain the fractional shift of the frequency of the demon:
\begin{align}
	\frac{\delta\omega}{\omega} =-\frac{9}{2} q^2 v_q \frac{D^{(3)}_h}{\omega^4} |E_{\omega,q}|^2 =-\frac{3}{4} (1-C_{\mathrm{ise}}) \xi
	\,,
	\label{eqn:kerr_shift}
\end{align}
where $\xi$ is defined by Eq.~\eqref{eqn:xi}.
The negative sign of the shift means the collective mode is softened by the strong field. The reason for this is
that the third-order conductivity is opposite in sign compared to the linear one, which is due to the current $\mathbf{j}$ being a concave function of the momentum density $\mathbf{p}$ in a Dirac fluid.
The results for the original and shifted demon dispersion in graphene is illustrated by 
Fig.~\ref{fig:Kerr}. It is remarkable that
an appreciable shift occurs already at a relatively low field of $E = 10^4 \unit{V/cm}$.

\section{Third order circular birefringence}
\label{sec:cd}
\begin{figure}[t]
	\includegraphics[width=3.0 in]{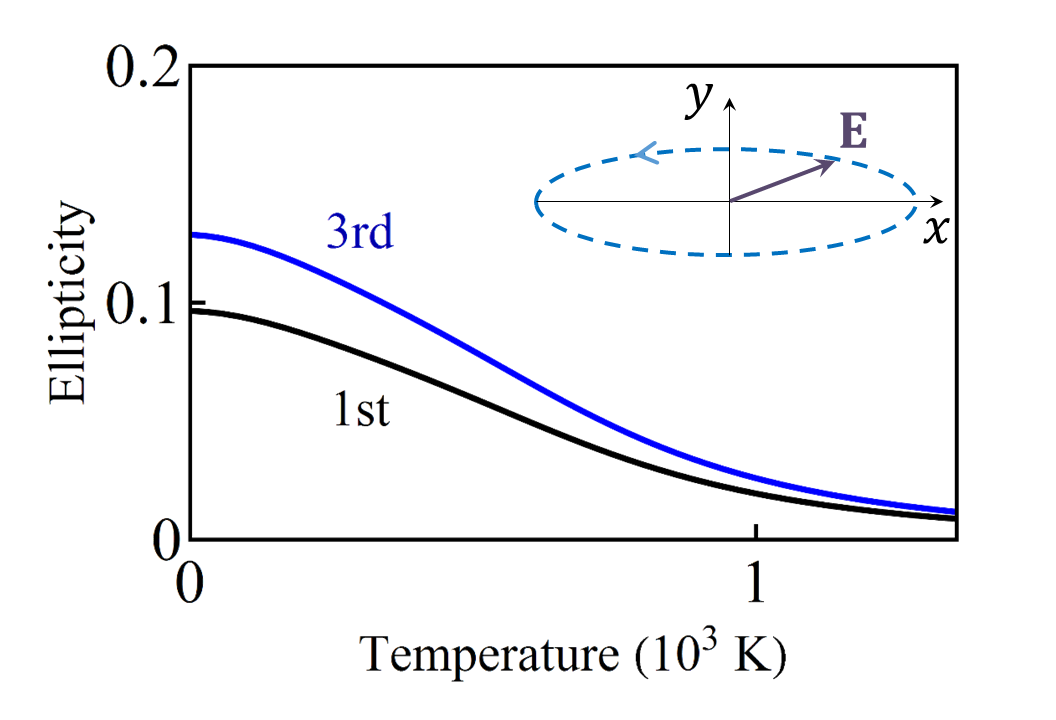}
	\caption{Circular birefrigence in graphene according to the hydrodynamic theory. The blue line is the ellipticity of the third-harmonic light ($|E_y/E_x|$) as a function of temperature at $n=10^{12} \unit{cm^{-2}}$. The black line is the ellipticity of the first-harmonic reflected light. The frequency is $\omega = 1.41 \unit{THz}$, the magnetic field is $B=0.1 \unit{T}$. Inset: illustration of the elliptical polarization.}
	\label{fig:circular_birefringence}
\end{figure}
In the presence of an applied magnetic field, there is a finite second term $\propto \omega_c$ in Eq.~\eqref{eq:sigma3_B}, which causes the third-order circular birefringence.
As in Sec.~\ref{sec:magnetic}, let us consider a 2D system subject to a normally incident monochromatic light of frequency $\omega$ with
the electric field polarized in the $x$-direction. The generated third-order current has frequency $\omega_s=3\omega$ and has a nonzero $y$-component
\begin{align}
j_x(3\omega) &= \sigma^{(3)}_{xxxx} E^3_x(\omega)=i D_h^{(3)} \frac{3}{\omega^3} \,, \notag \\
j_y(3\omega) &= \sigma^{(3)}_{yxxx} E^3_x(\omega)=i D_h^{(3)} \frac{3}{\omega^3}
\left(\frac{4 i\omega_c}{3 \omega}\right)  \,.
\label{eq:jx_jy}
\end{align}
In the dissipationless limit, $\omega$ is real, and $j_y$ has a $\pi/2$ phase difference relative to $j_x$. Therefore, the third harmonic light will be elliptically polarized with the principal axis along $x$. Its
ellipticity, conventionally denoted by $\tan \theta$, is given by
\begin{align}
\tan \theta = \left|\frac{j_y}{j_x} \right| = \frac{4}{3} \frac{\omega_c}{\omega} \,.
\label{eq:ellipticity}
\end{align}
Therefore, the ellipticity scales as $\omega_c = e B / m^{\ast} c$, which decays with temperature if the carrier density $n$ is fixed.
This is illustrated by Fig.~\ref{fig:circular_birefringence} for the case of graphene.
From Eq.~\eqref{Eq_sigma}, there is also circular birefringence in the linear response,
with $\tan \theta = \omega_c/\omega$, which differs only by the constant numerical factor $4 / 3$. It is also plotted in Fig.~\ref{fig:circular_birefringence},
for an easy comparison.

\section{Discussion}
\label{sec:discussion}

We showed that the third-order nonlinear conductivity $\sigma^{(3)}$ of a Dirac fluid has a universal functional form for any mass, chemical potential, temperature, and space dimension.
It is remarkable that the third-order and the linear conductivities are simply related through Eqs.~\eqref{eqn:D_h} and \eqref{eqn:sigma3_sigma1}.
Although we have used graphene as an example in the numerical calculations, our formulas, e.g., Eqs.~\eqref{eqn:sigma_3} and \eqref{eqn:D_h}, hold for any Lorentz-invariant Dirac fluid.
As such, these formulas should be a good approximation
to surface states of topological insulators and Dirac/Weyl semimetals, provided they are
in the hydrodynamic regime. 
We also studied the field-induced renormalization of the dispersion of the collective modes (demons) and
the third-order circular birefringence in the presence of a static magnetic field.

In the future it would be interesting to investigate hydrodynamics
of non-Dirac fluids, that is, systems without 
the Lorentz symmetry.
This will be important for more realistic modeling of ultrapure solid-state systems where hydrodynamic regime has been reported (GaAs, graphene, and $\mathrm{PdCoO_2}$). In the above systems, although the quasi particle band dispersion is approximately Dirac-like, the Coulomb interaction tends to break this quasi Lorentz symmetry because it propagates with the speed of light $c$ rather than $v$. It would also be interesting to study nonlinear thermal transport in the Dirac fluid. The case of phonon fluids has been studied half a century ago\cite{SHKLOSVKII.1969}.

\acknowledgments
This work is supported
% by the Department of Energy under Grant DE-SC0012592,
by the Office of Naval Research under Grant N00014-15-1-2671,
by the National Science Foundation under Grant ECCS-1640173,
and by the Semiconductor Research Corporation (SRC)
through the SRC-NRI Center for Excitonic Devices, research 2701.002.
D.~N.~B. is an investigator in Quantum Materials funded by
the Gordon and Betty Moore Foundation's EPiQS Initiative
through Grant No.~GBMF4533. We thank B. I. Shklovskii for helpful discussions.

%%%%%%%%%%%%%%%%%%%%%%%%%%%%%%%%%%%%%%%%%%%%%%%%%%%%%%%%%%%%%%%%%%%%%%%%%%%%%%%%%%
%%%%%%%%%%%%%%%%%%%%%%%%%%%%%%%%%%%%%%%%%%%%%%%%%%%%%%%%%%%%%%%%%%%%%%%%%%%%%%%%%%
%%%%%%%%%%%%%%%%%%%%%%%%%%%%%%%%%%%%%%%%%%%%%%%%%%%%%%%%%%%%%%%%%%%%%%%%%%%%%%%%%%
\appendix
\begin{widetext}

\section{Thermodynamic quantities in graphene}	
\label{appendix:thermodynamics}

For convenience, below we list the expressions for the thermodynamic quantities of graphene in the noninteracting limit. The charge density $n(\mu,T)$ \cite{Sun2016} is
\begin{align}
n &= \int\limits_{-\infty}^{\infty} \left[ f(\mu,T,\epsilon) - f(0, 0,\epsilon) \right] g(\epsilon) d\epsilon
= \frac{1}{\pi} \frac{\mu^2}{\hbar^2 v_F^2}  \Bigl[1+ \frac{\pi^2}{3}
\frac{T^2}{\mu^2}
+ 4\, \frac{T^2}{\mu^2} \mathrm{Li}_2\bigl(-e^{-\mu/ T}\bigl) \Bigr] \,
\label{eqn:n}
\end{align}
where $\mathrm{Li}_s(x)$ is the polylogarithm function. The energy density $n_E$ is defined relative to the $(\mu,T)=(0,0)$ case:
\begin{align}
n_E &= \int\limits_{-\infty}^{\infty} \left[ f(\mu,T,\epsilon) - f(0, 0,\epsilon)\right] \epsilon g(\epsilon) d\epsilon
=\frac{2}{\pi} \frac{T^3}{\hbar^2 v_F^2} \Bigl[
\frac{\pi^2}{3}\, \frac{\mu}{T} + \frac{1}{3}\, \frac{\mu^3}{T^3} 
-4\, \mathrm{Li}_3 \bigl(-e^{-\mu/T}\bigr) \Bigr] \notag\\
&=\frac{2}{3\pi} \frac{\mu^3}{\hbar^2 v_F^2} \Bigl[1 +\pi^2 \frac{T^2}{\mu^2} 
-12 \frac{T^3}{\mu^3} \mathrm{Li}_3 \bigl(-e^{-\mu/T}\bigr) \Bigr] \,.
\label{eqn:n_E}
\end{align}
The enthalpy density is $W=\frac{3}{2} n_E$, the pressure is $P=\frac{1}{2} n_E$,
and the entropy density is  
\begin{align}
s = \left(\frac{\partial P}{\partial T}\right)_\mu
= \frac{1}{3\pi} \frac{\mu^2}{\hbar^2 v_F^2} \Bigl[2\pi^2 \frac{T}{\mu} -12 \frac{T}{\mu} \mathrm{Li}_2 \bigl(-e^{-\mu/T}\bigr)
- 36 \frac{T^2}{\mu^2} \mathrm{Li}_3 \bigl(-e^{-\mu/T}\bigr) \Bigr] \,.
\end{align}
The hydrodynamic effective mass $m^{\ast} (\mu,T)$ is
\begin{align}
m^{\ast} (\mu,T) = \frac{1}{ v^2} \frac{W(\mu,T)}{n(\mu,T)} \,.
\end{align}
The dimensionless bulk isentropic modulus is
\begin{align}
C_{\mathrm{ise}}=
= \frac{1}{m^{\ast}\! v^2}
\left( \frac{\partial P}{\partial n} \right)_{\mathrm{ise}}=\frac{1}{d} =\frac{1}{2} \,.
\end{align}

\section{Derivation of $\sigma^{(3)}$ with momentum and energy relaxation}
\label{appendix:derivation}
In the homogeneous case, the hydrodynamic equations are
\begin{align}
(\partial_t + \Gamma_d) p_i = \rho E_i ,\quad	\partial_t n_{E}   = \rho\, u_j E_j - \Gamma_E \delta n_E - \Gamma_k Wu^2 ,\, \quad
\partial_t \rho   =0 \,,
\label{eqn:charge_c2}
\end{align}
where
$\Gamma_d$ is the phenomenological momentum relaxation rate,
$\Gamma_E$ can be called the cooling rate,
$\delta n_E = n_E -  n_{Eeq}$ is the fluctuation of energy density with respect to its steady-state value,
and $\Gamma_k$ is the relaxation rate of the center-of-mass kinetic energy of a moving fluid.
%%
%And the dissipative case is
%%%
%\begin{align}
%	(\partial_t + \Gamma) u_i = \frac{1}{\gamma^2 W} \left( n E_i - u_i \partial_t P - u_i j_j E_j
%	\right) \,,
%\label{eqn:u3}
%\end{align}
%%%
The last equation entails $\rho$ is constant. Therefore $j_i^{(3)} = \rho u^{(3)}_i$ and the momentum $p_i$ is strictly linear in electric field, same as the dissipationless case. 
The third-order velocity can be found from Eq.~\eqref{Eq_third_order_simple}
\begin{align}
u_i^{(3)}  = - \left( W^{(2)}/W + {u^{(1)}}^2 \right) u_i^{(1)}
\label{Eq_u_3_1} \,.
\end{align}
%	%%
%	\begin{align}
%	\left(\partial_t + \Gamma_d\right) u^{(3)}_i = \frac{1}{ W} \left(  -\frac{\rho}{ W} \left(\gamma^2 W\right)^{(2)} E_i -  u^{(1)}_i \partial_t P^{(2)} - \rho u^{(1)}_i u^{(1)}_j E_j +  \Gamma_E u^{(1)}_i n_E^{(2)}  \right) \,,
%	\label{eqn:u3}
%	\end{align}
%	%%
By rotational symmetry, the leading order perturbation to the scalar quantities are second order in the electric field \cite{Sun2017}
    \begin{equation}
    \begin{split}
    n_0^{(2)} = -\frac{1}{2}n_0 {u^{(1)}}^2 \,,  \quad
    n_E^{(2)} = \frac{\omega_2^+ - i\Gamma_k}{\omega_1+\omega_2+i\Gamma_E} W u^{(1)}_{1i} u^{(1)}_{2i} + \mathrm{perm} \,, \quad
    n_{E0}^{(2)} = n^{(2)}_E - W {u^{(1)}}^2 \,.
%    \\
%    P^{(2)} &=  n^{(2)}_0 \frac{\partial P}{\partial n_0}
%    + n^{(2)}_{E0} \frac{\partial P}{\partial n_{E0}}
%    = \frac{1}{2} \left(-C_{\mathrm{ise}}
%    + \frac{2 i \Gamma_{d}-i\Gamma_E}{\omega_1+\omega_2+i\Gamma_E}
%    \frac{\partial P}{\partial n_{E0}}\right)
%    W u^{(1)}_{1i} u^{(1)}_{2i} + (1 \leftrightarrow 2)   \,.
    \label{eqn:second}
    \end{split}
    \end{equation}
where ``$\mathrm{perm}$'' stands for permutations
among subscripts $1$, $2$, and $3$, corresponding to
frequencies $\omega_1$, $\omega_2$, and $\omega_3$,
respectively. Note the difference between density $n_0$ and energy density $n_{E0}$ in proper frame and their counterparts $n$, $n_E $ in lab frame. With the second order expansion of the enthalpy
	\begin{align}
	 W^{(2)} = \left(\frac{\partial W}{\partial n_0} \right)_{n_{E0}} n_0^{(2)} +\left(\frac{\partial W}{\partial n_{E0}} \right)_{n_{0}} n_{E0}^{(2)} \,
	\end{align}
we are ready to write down the third order flow velocity
\begin{align}
u_i^{(3)}  = - \left[ \frac{1}{W}\left(\frac{\partial W}{\partial n_0} \right)_{n_{E0}}  \left(-\frac{1}{2}n_0 \right) {u^{(1)}}^2 + \left(\frac{\partial W}{\partial n_{E0}} \right)_{n_{0}} \left(  \frac{\omega_2^+ - i\Gamma_k}{\omega_1+\omega_2+i\Gamma_E}  u^{(1)}_{1j} u^{(1)}_{2j} + \mathrm{perm}\right) - \left(\frac{\partial W}{\partial n_{E0}} \right)_{n_{0}} {u^{(1)}}^2 +{u^{(1)}}^2 \right]  u_i^{(1)}
\label{Eq_u_3_1} \,.
\end{align}
where we defined $\omega_a^{+} \equiv \omega_a+i\Gamma_d $. 
The equation for the Fourier amplitude $u^{(3)}(\omega_s)$ of the combined
frequency $\omega_s = \omega_1 + \omega_2 + \omega_3$ becomes
\begin{align}
u_i^{(3)}  &= - \left[ \frac{1}{W}\left(\frac{\partial W}{\partial n_0} \right)_{n_{E0}}  \left(-\frac{1}{2}n_0 \right)  + \left(\frac{\partial W}{\partial n_{E0}} \right)_{n_{0}} \left(  \frac{\omega_2^+ - i\Gamma_k}{\omega_1+\omega_2+i\Gamma_E}  \right) - \left(\frac{\partial W}{\partial n_{E0}} \right)_{n_{0}}  + 1 \right]   u^{(1)}_{1j} u^{(1)}_{2j} u^{(1)}_{3i} + \mathrm{perm} \notag\\
&=  \left[ \frac{1}{2} \frac{n_0}{W}\left(\frac{\partial W}{\partial n_0} \right)_{n_{E0}} + \left(\frac{\partial W}{\partial n_{E0}} \right)_{n_{0}}  - 1  - \frac{1}{2}\left(\frac{\partial W}{\partial n_{E0}} \right)_{n_{0}} \left(  \frac{\omega_1^+ + \omega_2^+ - 2i\Gamma_k}{\omega_1+\omega_2+i\Gamma_E}  \right)  \right]   u^{(1)}_{1j} u^{(1)}_{2j} u^{(1)}_{3i} + \mathrm{perm} \notag \\
&=   \frac{1}{2}\left[ C_{\mathrm{ise}} - 1 - \left(\frac{\partial W}{\partial n_{E0}} \right)_{n_{0}} \left(  \frac{2i\Gamma_d - i\Gamma_E- 2i\Gamma_k}{\omega_1+\omega_2+i\Gamma_E}  \right)  \right]   u^{(1)}_{1j} u^{(1)}_{2j} u^{(1)}_{3i} + \mathrm{perm} 
\label{eqn:u3_dirty} \,.
\end{align}

After the standard symmetrization procedure, Eq.~\eqref{eqn:u3_dirty} renders $\sigma^{(3)}$ with dissipation. The cooling rate $\Gamma_E$ could arise due to electron-phonon coupling and is crucial for eliminating the divergence of $\sigma^{(3)}$ in the dc limit. In this dc limit, due to work done by the electric field, the electron-hole fluid would be heated up by order $E^2/\Gamma_E$, thus inducing a large correction to the current at the third order. This is a physical reason why
setting $\Gamma_E \to 0$ would lead to a diverging $\sigma^{(3)}$.
	
In the dissipationless limit, Eq.~\eqref{eqn:u3_dirty} becomes Eq.~\eqref{Eq_third_order}. Moreover, it can be readily checked that if $\Gamma_E+2\Gamma_k=2\Gamma_d$, Eq.~\eqref{eqn:u3_dirty} becomes identical to Eq.~\eqref{Eq_third_order} as well. Under this condition, the fluid dynamics becomes isentropic again: the kinetic energy is lost to the environment due to momentum relaxation instead of converted into heat of the fluid. 
%	Indeed, the kinetic energy is quadratic in momentum $\delta n_E = ap^2 + O(p^4)$, thus $\partial_t \delta n_E \approx 2ap\partial_t p = 2ap (-\Gamma_d) p =-2\Gamma_d \delta n_E $ .
	%%
\end{widetext}

\section{Relativistic Boltzmann equation}
\label{appendix:RBE}
The relativistic Boltzmann equation is
\begin{align}
	\left( P^\mu \partial_\mu + F_{\mu\nu} P^\mu \partial_{P_\nu} \right) f_R (X,P) =I[f_R] \,,
	\label{Eq_rela_boltzmann}
\end{align}
where $f_R (X,P)$ is the relativistic distribution function and $I[f_R]$ is the collision integral due to interactions. 
Note that the space-time coordinate $X^{\mu}$ and the momentum $P^{\mu}=mu^{\mu}$ are covariant ones. For a given $X^{\mu}$, the distribution function $f_R (X,P)$ can be defined as the density of world lines whose local tangent is $P^{\mu}$. Mathematically, $f_R (X,P)$ is a scalar function defined on the tangent bundle of the $d+1$ dimensional space-time. If we focus on one species of particle with a fixed mass $m$, then $f_R (X,P)$ is related to the ordinary distribution function through
\begin{align}
	f_R (X,P) = f(t,\mathbf{r},\mathbf{p}) \delta(E^2 - p^2 - m^2) \,.
	\label{Eq_relation_ordinary}
\end{align}
We integrate Eq.~\eqref{Eq_rela_boltzmann} over $P$ to get the charge continuity equation:
\begin{align}
	&\partial_\mu J^\mu =0 \,, \notag\\
	& j^\mu = \int d P P^\mu f_R (X,P) =\int d\mathbf{p} \; (1, \mathbf{v}) f(t,\mathbf{r},\mathbf{p})  \,,
	\label{Eq_conti}
\end{align}
which is the second equation in Eq.~\eqref{Eq_hydro1}.

Next, we multiply Eq.~\eqref{Eq_rela_boltzmann} by $P^\alpha$ and again integrate it over $P$.
We get the continuity equation for the energy-momentum tensor,
\begin{align}
	&\partial_\mu T^{\mu\nu} = F_\mu^\nu J^\mu \,, \notag\\
	& T^{\mu\nu} = \int d P P^\mu P^\nu f_R (X,P) =\int d\mathbf{p} \; P^\mu P^\nu \frac{1}{E} f(t,\mathbf{r},\mathbf{p})  \,,
	\label{Eq_stress}
\end{align} 
which is
the first equation in Eq.~\eqref{Eq_hydro1}.

%%%%%%%%%%%%%%%%%%%%%%%%%%%%%%%%%%%%%%%%%%%%%
\bibliographystyle{apsrev4-1}
\bibliography{DiracFluid_library}

\begin{thebibliography}{40}%
\makeatletter
\providecommand \@ifxundefined [1]{%
 \@ifx{#1\undefined}
}%
\providecommand \@ifnum [1]{%
 \ifnum #1\expandafter \@firstoftwo
 \else \expandafter \@secondoftwo
 \fi
}%
\providecommand \@ifx [1]{%
 \ifx #1\expandafter \@firstoftwo
 \else \expandafter \@secondoftwo
 \fi
}%
\providecommand \natexlab [1]{#1}%
\providecommand \enquote  [1]{``#1''}%
\providecommand \bibnamefont  [1]{#1}%
\providecommand \bibfnamefont [1]{#1}%
\providecommand \citenamefont [1]{#1}%
\providecommand \href@noop [0]{\@secondoftwo}%
\providecommand \href [0]{\begingroup \@sanitize@url \@href}%
\providecommand \@href[1]{\@@startlink{#1}\@@href}%
\providecommand \@@href[1]{\endgroup#1\@@endlink}%
\providecommand \@sanitize@url [0]{\catcode `\\12\catcode `\$12\catcode
  `\&12\catcode `\#12\catcode `\^12\catcode `\_12\catcode `\%12\relax}%
\providecommand \@@startlink[1]{}%
\providecommand \@@endlink[0]{}%
\providecommand \url  [0]{\begingroup\@sanitize@url \@url }%
\providecommand \@url [1]{\endgroup\@href {#1}{\urlprefix }}%
\providecommand \urlprefix  [0]{URL }%
\providecommand \Eprint [0]{\href }%
\providecommand \doibase [0]{http://dx.doi.org/}%
\providecommand \selectlanguage [0]{\@gobble}%
\providecommand \bibinfo  [0]{\@secondoftwo}%
\providecommand \bibfield  [0]{\@secondoftwo}%
\providecommand \translation [1]{[#1]}%
\providecommand \BibitemOpen [0]{}%
\providecommand \bibitemStop [0]{}%
\providecommand \bibitemNoStop [0]{.\EOS\space}%
\providecommand \EOS [0]{\spacefactor3000\relax}%
\providecommand \BibitemShut  [1]{\csname bibitem#1\endcsname}%
\let\auto@bib@innerbib\@empty
%</preamble>
\bibitem [{\citenamefont {Gurzhi}(1968)}]{Gurzhi1968}%
  \BibitemOpen
  \bibfield  {author} {\bibinfo {author} {\bibfnamefont {R.~N.}\ \bibnamefont
  {Gurzhi}},\ }\href {\doibase 10.1070/PU1968v011n02ABEH003815} {\bibfield
  {journal} {\bibinfo  {journal} {Sov. Phys. Uspekhi}\ }\textbf {\bibinfo
  {volume} {11}},\ \bibinfo {pages} {255} (\bibinfo {year} {1968})}\BibitemShut
  {NoStop}%
\bibitem [{\citenamefont {Andreev}\ \emph {et~al.}(2011)\citenamefont
  {Andreev}, \citenamefont {Kivelson},\ and\ \citenamefont
  {Spivak}}]{Andreev2011}%
  \BibitemOpen
  \bibfield  {author} {\bibinfo {author} {\bibfnamefont {A.~V.}\ \bibnamefont
  {Andreev}}, \bibinfo {author} {\bibfnamefont {S.~A.}\ \bibnamefont
  {Kivelson}}, \ and\ \bibinfo {author} {\bibfnamefont {B.}~\bibnamefont
  {Spivak}},\ }\href {\doibase 10.1103/PhysRevLett.106.256804} {\bibfield
  {journal} {\bibinfo  {journal} {Phys. Rev. Lett.}\ }\textbf {\bibinfo
  {volume} {106}},\ \bibinfo {pages} {256804} (\bibinfo {year}
  {2011})}\BibitemShut {NoStop}%
\bibitem [{\citenamefont {Landau}\ and\ \citenamefont
  {Lifshitz}(1987)}]{Landau.6}%
  \BibitemOpen
  \bibfield  {author} {\bibinfo {author} {\bibfnamefont {L.~D.}\ \bibnamefont
  {Landau}}\ and\ \bibinfo {author} {\bibfnamefont {E.~M.}\ \bibnamefont
  {Lifshitz}},\ }\href@noop {} {\emph {\bibinfo {title} {{Fluid Mechanics}}}},\
  \bibinfo {edition} {2nd}\ ed.\ (\bibinfo  {publisher} {Pergamon Press},\
  \bibinfo {address} {Oxford},\ \bibinfo {year} {1987})\BibitemShut {NoStop}%
\bibitem [{\citenamefont {de~Jong}\ and\ \citenamefont
  {Molenkamp}(1995)}]{DeJong1995}%
  \BibitemOpen
  \bibfield  {author} {\bibinfo {author} {\bibfnamefont {M.~J.~M.}\
  \bibnamefont {de~Jong}}\ and\ \bibinfo {author} {\bibfnamefont {L.~W.}\
  \bibnamefont {Molenkamp}},\ }\href {\doibase 10.1103/PhysRevB.51.13389}
  {\bibfield  {journal} {\bibinfo  {journal} {Phys. Rev. B}\ }\textbf {\bibinfo
  {volume} {51}},\ \bibinfo {pages} {13389} (\bibinfo {year}
  {1995})}\BibitemShut {NoStop}%
\bibitem [{\citenamefont {Bandurin}\ \emph {et~al.}(2016)\citenamefont
  {Bandurin}, \citenamefont {Torre}, \citenamefont {Kumar}, \citenamefont {{Ben
  Shalom}}, \citenamefont {Tomadin}, \citenamefont {Principi}, \citenamefont
  {Auton}, \citenamefont {Khestanova}, \citenamefont {Novoselov}, \citenamefont
  {Grigorieva}, \citenamefont {Ponomarenko}, \citenamefont {Geim},\ and\
  \citenamefont {Polini}}]{Bandurin2016}%
  \BibitemOpen
  \bibfield  {author} {\bibinfo {author} {\bibfnamefont {D.~A.}\ \bibnamefont
  {Bandurin}}, \bibinfo {author} {\bibfnamefont {I.}~\bibnamefont {Torre}},
  \bibinfo {author} {\bibfnamefont {R.~K.}\ \bibnamefont {Kumar}}, \bibinfo
  {author} {\bibfnamefont {M.}~\bibnamefont {{Ben Shalom}}}, \bibinfo {author}
  {\bibfnamefont {A.}~\bibnamefont {Tomadin}}, \bibinfo {author} {\bibfnamefont
  {A.}~\bibnamefont {Principi}}, \bibinfo {author} {\bibfnamefont {G.~H.}\
  \bibnamefont {Auton}}, \bibinfo {author} {\bibfnamefont {E.}~\bibnamefont
  {Khestanova}}, \bibinfo {author} {\bibfnamefont {K.~S.}\ \bibnamefont
  {Novoselov}}, \bibinfo {author} {\bibfnamefont {I.~V.}\ \bibnamefont
  {Grigorieva}}, \bibinfo {author} {\bibfnamefont {L.~A.}\ \bibnamefont
  {Ponomarenko}}, \bibinfo {author} {\bibfnamefont {A.~K.}\ \bibnamefont
  {Geim}}, \ and\ \bibinfo {author} {\bibfnamefont {M.}~\bibnamefont
  {Polini}},\ }\href {\doibase 10.1126/science.aad0201} {\bibfield  {journal}
  {\bibinfo  {journal} {Science}\ }\textbf {\bibinfo {volume} {351}},\ \bibinfo
  {pages} {1055} (\bibinfo {year} {2016})}\BibitemShut {NoStop}%
\bibitem [{\citenamefont {Crossno}\ \emph {et~al.}(2016)\citenamefont
  {Crossno}, \citenamefont {Shi}, \citenamefont {Wang}, \citenamefont {Liu},
  \citenamefont {Harzheim}, \citenamefont {Lucas}, \citenamefont {Sachdev},
  \citenamefont {Kim}, \citenamefont {Taniguchi}, \citenamefont {Watanabe},
  \citenamefont {Ohki},\ and\ \citenamefont {Fong}}]{Crossno2016}%
  \BibitemOpen
  \bibfield  {author} {\bibinfo {author} {\bibfnamefont {J.}~\bibnamefont
  {Crossno}}, \bibinfo {author} {\bibfnamefont {J.~K.}\ \bibnamefont {Shi}},
  \bibinfo {author} {\bibfnamefont {K.}~\bibnamefont {Wang}}, \bibinfo {author}
  {\bibfnamefont {X.}~\bibnamefont {Liu}}, \bibinfo {author} {\bibfnamefont
  {A.}~\bibnamefont {Harzheim}}, \bibinfo {author} {\bibfnamefont
  {A.}~\bibnamefont {Lucas}}, \bibinfo {author} {\bibfnamefont
  {S.}~\bibnamefont {Sachdev}}, \bibinfo {author} {\bibfnamefont
  {P.}~\bibnamefont {Kim}}, \bibinfo {author} {\bibfnamefont {T.}~\bibnamefont
  {Taniguchi}}, \bibinfo {author} {\bibfnamefont {K.}~\bibnamefont {Watanabe}},
  \bibinfo {author} {\bibfnamefont {T.~A.}\ \bibnamefont {Ohki}}, \ and\
  \bibinfo {author} {\bibfnamefont {K.~C.}\ \bibnamefont {Fong}},\ }\href
  {\doibase 10.1126/science.aad0343} {\bibfield  {journal} {\bibinfo  {journal}
  {Science}\ }\textbf {\bibinfo {volume} {351}},\ \bibinfo {pages} {1058}
  (\bibinfo {year} {2016})}\BibitemShut {NoStop}%
\bibitem [{\citenamefont {{Krishna Kumar}}\ \emph {et~al.}(2017)\citenamefont
  {{Krishna Kumar}}, \citenamefont {Bandurin}, \citenamefont {Pellegrino},
  \citenamefont {Cao}, \citenamefont {Principi}, \citenamefont {Guo},
  \citenamefont {Auton}, \citenamefont {{Ben Shalom}}, \citenamefont
  {Ponomarenko}, \citenamefont {Falkovich}, \citenamefont {Watanabe},
  \citenamefont {Taniguchi}, \citenamefont {Grigorieva}, \citenamefont
  {Levitov}, \citenamefont {Polini},\ and\ \citenamefont {Geim}}]{Kumar2017}%
  \BibitemOpen
  \bibfield  {author} {\bibinfo {author} {\bibfnamefont {R.}~\bibnamefont
  {{Krishna Kumar}}}, \bibinfo {author} {\bibfnamefont {D.~A.}\ \bibnamefont
  {Bandurin}}, \bibinfo {author} {\bibfnamefont {F.~M.~D.}\ \bibnamefont
  {Pellegrino}}, \bibinfo {author} {\bibfnamefont {Y.}~\bibnamefont {Cao}},
  \bibinfo {author} {\bibfnamefont {A.}~\bibnamefont {Principi}}, \bibinfo
  {author} {\bibfnamefont {H.}~\bibnamefont {Guo}}, \bibinfo {author}
  {\bibfnamefont {G.~H.}\ \bibnamefont {Auton}}, \bibinfo {author}
  {\bibfnamefont {M.}~\bibnamefont {{Ben Shalom}}}, \bibinfo {author}
  {\bibfnamefont {L.~A.}\ \bibnamefont {Ponomarenko}}, \bibinfo {author}
  {\bibfnamefont {G.}~\bibnamefont {Falkovich}}, \bibinfo {author}
  {\bibfnamefont {K.}~\bibnamefont {Watanabe}}, \bibinfo {author}
  {\bibfnamefont {T.}~\bibnamefont {Taniguchi}}, \bibinfo {author}
  {\bibfnamefont {I.~V.}\ \bibnamefont {Grigorieva}}, \bibinfo {author}
  {\bibfnamefont {L.~S.}\ \bibnamefont {Levitov}}, \bibinfo {author}
  {\bibfnamefont {M.}~\bibnamefont {Polini}}, \ and\ \bibinfo {author}
  {\bibfnamefont {A.~K.}\ \bibnamefont {Geim}},\ }\href {\doibase
  10.1038/nphys4240} {\bibfield  {journal} {\bibinfo  {journal} {Nat. Phys.}\ }
  (\bibinfo {year} {2017}),\ 10.1038/nphys4240}\BibitemShut {NoStop}%
\bibitem [{\citenamefont {Moll}\ \emph {et~al.}(2016)\citenamefont {Moll},
  \citenamefont {Kushwaha}, \citenamefont {Nandi}, \citenamefont {Schmidt},\
  and\ \citenamefont {Mackenzie}}]{Moll2016}%
  \BibitemOpen
  \bibfield  {author} {\bibinfo {author} {\bibfnamefont {P.~J.~W.}\
  \bibnamefont {Moll}}, \bibinfo {author} {\bibfnamefont {P.}~\bibnamefont
  {Kushwaha}}, \bibinfo {author} {\bibfnamefont {N.}~\bibnamefont {Nandi}},
  \bibinfo {author} {\bibfnamefont {B.}~\bibnamefont {Schmidt}}, \ and\
  \bibinfo {author} {\bibfnamefont {A.~P.}\ \bibnamefont {Mackenzie}},\ }\href
  {\doibase 10.1126/science.aac8385} {\bibfield  {journal} {\bibinfo  {journal}
  {Science}\ }\textbf {\bibinfo {volume} {351}},\ \bibinfo {pages} {1061}
  (\bibinfo {year} {2016})}\BibitemShut {NoStop}%
\bibitem [{\citenamefont {M{\"{u}}ller}\ \emph {et~al.}(2008)\citenamefont
  {M{\"{u}}ller}, \citenamefont {Fritz},\ and\ \citenamefont
  {Sachdev}}]{Muller2008a}%
  \BibitemOpen
  \bibfield  {author} {\bibinfo {author} {\bibfnamefont {M.}~\bibnamefont
  {M{\"{u}}ller}}, \bibinfo {author} {\bibfnamefont {L.}~\bibnamefont {Fritz}},
  \ and\ \bibinfo {author} {\bibfnamefont {S.}~\bibnamefont {Sachdev}},\ }\href
  {\doibase 10.1103/PhysRevB.78.115406} {\bibfield  {journal} {\bibinfo
  {journal} {Phys. Rev. B}\ }\textbf {\bibinfo {volume} {78}},\ \bibinfo
  {pages} {115406} (\bibinfo {year} {2008})}\BibitemShut {NoStop}%
\bibitem [{\citenamefont {M{\"{u}}ller}\ \emph {et~al.}(2009)\citenamefont
  {M{\"{u}}ller}, \citenamefont {Schmalian},\ and\ \citenamefont
  {Fritz}}]{Muller2009}%
  \BibitemOpen
  \bibfield  {author} {\bibinfo {author} {\bibfnamefont {M.}~\bibnamefont
  {M{\"{u}}ller}}, \bibinfo {author} {\bibfnamefont {J.}~\bibnamefont
  {Schmalian}}, \ and\ \bibinfo {author} {\bibfnamefont {L.}~\bibnamefont
  {Fritz}},\ }\href {\doibase 10.1103/PhysRevLett.103.025301} {\bibfield
  {journal} {\bibinfo  {journal} {Phys. Rev. Lett.}\ }\textbf {\bibinfo
  {volume} {103}},\ \bibinfo {pages} {025301} (\bibinfo {year}
  {2009})}\BibitemShut {NoStop}%
\bibitem [{\citenamefont {Phan}\ \emph {et~al.}()\citenamefont {Phan},
  \citenamefont {Song},\ and\ \citenamefont {Levitov}}]{Phan2013}%
  \BibitemOpen
  \bibfield  {author} {\bibinfo {author} {\bibfnamefont {T.~V.}\ \bibnamefont
  {Phan}}, \bibinfo {author} {\bibfnamefont {J.~C.~W.}\ \bibnamefont {Song}}, \
  and\ \bibinfo {author} {\bibfnamefont {L.~S.}\ \bibnamefont {Levitov}},\
  }\href {http://arxiv.org/abs/1306.4972} {\enquote {\bibinfo {title}
  {{Ballistic Heat Transfer and Energy Waves in an Electron System}},}\
  }\bibinfo {note} {(unpublished)},\ \Eprint {http://arxiv.org/abs/1306.4972}
  {arXiv:1306.4972} \BibitemShut {NoStop}%
\bibitem [{\citenamefont {Forcella}\ \emph {et~al.}(2014)\citenamefont
  {Forcella}, \citenamefont {Zaanen}, \citenamefont {Valentinis},\ and\
  \citenamefont {van~der Marel}}]{Forcella2014}%
  \BibitemOpen
  \bibfield  {author} {\bibinfo {author} {\bibfnamefont {D.}~\bibnamefont
  {Forcella}}, \bibinfo {author} {\bibfnamefont {J.}~\bibnamefont {Zaanen}},
  \bibinfo {author} {\bibfnamefont {D.}~\bibnamefont {Valentinis}}, \ and\
  \bibinfo {author} {\bibfnamefont {D.}~\bibnamefont {van~der Marel}},\ }\href
  {\doibase 10.1103/PhysRevB.90.035143} {\bibfield  {journal} {\bibinfo
  {journal} {Phys. Rev. B}\ }\textbf {\bibinfo {volume} {90}},\ \bibinfo
  {pages} {035143} (\bibinfo {year} {2014})}\BibitemShut {NoStop}%
\bibitem [{\citenamefont {Briskot}\ \emph {et~al.}(2015)\citenamefont
  {Briskot}, \citenamefont {Sch{\"{u}}tt}, \citenamefont {Gornyi},
  \citenamefont {Titov}, \citenamefont {Narozhny},\ and\ \citenamefont
  {Mirlin}}]{Briskot2015}%
  \BibitemOpen
  \bibfield  {author} {\bibinfo {author} {\bibfnamefont {U.}~\bibnamefont
  {Briskot}}, \bibinfo {author} {\bibfnamefont {M.}~\bibnamefont
  {Sch{\"{u}}tt}}, \bibinfo {author} {\bibfnamefont {I.~V.}\ \bibnamefont
  {Gornyi}}, \bibinfo {author} {\bibfnamefont {M.}~\bibnamefont {Titov}},
  \bibinfo {author} {\bibfnamefont {B.~N.}\ \bibnamefont {Narozhny}}, \ and\
  \bibinfo {author} {\bibfnamefont {A.~D.}\ \bibnamefont {Mirlin}},\ }\href
  {\doibase 10.1103/PhysRevB.92.115426} {\bibfield  {journal} {\bibinfo
  {journal} {Phys. Rev. B}\ }\textbf {\bibinfo {volume} {92}},\ \bibinfo
  {pages} {115426} (\bibinfo {year} {2015})}\BibitemShut {NoStop}%
\bibitem [{\citenamefont {Narozhny}\ \emph {et~al.}(2015)\citenamefont
  {Narozhny}, \citenamefont {Gornyi}, \citenamefont {Titov}, \citenamefont
  {Sch{\"{u}}tt},\ and\ \citenamefont {Mirlin}}]{Narozhny2015}%
  \BibitemOpen
  \bibfield  {author} {\bibinfo {author} {\bibfnamefont {B.~N.}\ \bibnamefont
  {Narozhny}}, \bibinfo {author} {\bibfnamefont {I.~V.}\ \bibnamefont
  {Gornyi}}, \bibinfo {author} {\bibfnamefont {M.}~\bibnamefont {Titov}},
  \bibinfo {author} {\bibfnamefont {M.}~\bibnamefont {Sch{\"{u}}tt}}, \ and\
  \bibinfo {author} {\bibfnamefont {A.~D.}\ \bibnamefont {Mirlin}},\ }\href
  {\doibase 10.1103/PhysRevB.91.035414} {\bibfield  {journal} {\bibinfo
  {journal} {Phys. Rev. B}\ }\textbf {\bibinfo {volume} {91}},\ \bibinfo
  {pages} {035414} (\bibinfo {year} {2015})}\BibitemShut {NoStop}%
\bibitem [{\citenamefont {Principi}\ and\ \citenamefont
  {Vignale}(2015{\natexlab{a}})}]{Principi2015b}%
  \BibitemOpen
  \bibfield  {author} {\bibinfo {author} {\bibfnamefont {A.}~\bibnamefont
  {Principi}}\ and\ \bibinfo {author} {\bibfnamefont {G.}~\bibnamefont
  {Vignale}},\ }\href {\doibase 10.1103/PhysRevB.91.205423} {\bibfield
  {journal} {\bibinfo  {journal} {Phys. Rev. B}\ }\textbf {\bibinfo {volume}
  {91}},\ \bibinfo {pages} {205423} (\bibinfo {year}
  {2015}{\natexlab{a}})}\BibitemShut {NoStop}%
\bibitem [{\citenamefont {Principi}\ and\ \citenamefont
  {Vignale}(2015{\natexlab{b}})}]{Principi2015}%
  \BibitemOpen
  \bibfield  {author} {\bibinfo {author} {\bibfnamefont {A.}~\bibnamefont
  {Principi}}\ and\ \bibinfo {author} {\bibfnamefont {G.}~\bibnamefont
  {Vignale}},\ }\href {\doibase 10.1103/PhysRevLett.115.056603} {\bibfield
  {journal} {\bibinfo  {journal} {Phys. Rev. Lett.}\ }\textbf {\bibinfo
  {volume} {115}},\ \bibinfo {pages} {056603} (\bibinfo {year}
  {2015}{\natexlab{b}})}\BibitemShut {NoStop}%
\bibitem [{\citenamefont {Sun}\ \emph {et~al.}(2016)\citenamefont {Sun},
  \citenamefont {Basov},\ and\ \citenamefont {Fogler}}]{Sun2016}%
  \BibitemOpen
  \bibfield  {author} {\bibinfo {author} {\bibfnamefont {Z.}~\bibnamefont
  {Sun}}, \bibinfo {author} {\bibfnamefont {D.~N.}\ \bibnamefont {Basov}}, \
  and\ \bibinfo {author} {\bibfnamefont {M.~M.}\ \bibnamefont {Fogler}},\
  }\href {\doibase 10.1103/PhysRevLett.117.076805} {\bibfield  {journal}
  {\bibinfo  {journal} {Phys. Rev. Lett.}\ }\textbf {\bibinfo {volume} {117}},\
  \bibinfo {pages} {076805} (\bibinfo {year} {2016})}\BibitemShut {NoStop}%
\bibitem [{\citenamefont {Lucas}\ \emph {et~al.}(2016)\citenamefont {Lucas},
  \citenamefont {Crossno}, \citenamefont {Fong}, \citenamefont {Kim},\ and\
  \citenamefont {Sachdev}}]{Lucas2016}%
  \BibitemOpen
  \bibfield  {author} {\bibinfo {author} {\bibfnamefont {A.}~\bibnamefont
  {Lucas}}, \bibinfo {author} {\bibfnamefont {J.}~\bibnamefont {Crossno}},
  \bibinfo {author} {\bibfnamefont {K.~C.}\ \bibnamefont {Fong}}, \bibinfo
  {author} {\bibfnamefont {P.}~\bibnamefont {Kim}}, \ and\ \bibinfo {author}
  {\bibfnamefont {S.}~\bibnamefont {Sachdev}},\ }\href {\doibase
  10.1103/PhysRevB.93.075426} {\bibfield  {journal} {\bibinfo  {journal} {Phys.
  Rev. B}\ }\textbf {\bibinfo {volume} {93}},\ \bibinfo {pages} {075426}
  (\bibinfo {year} {2016})}\BibitemShut {NoStop}%
\bibitem [{\citenamefont {Lucas}(2017)}]{Lucas2016b}%
  \BibitemOpen
  \bibfield  {author} {\bibinfo {author} {\bibfnamefont {A.}~\bibnamefont
  {Lucas}},\ }\href {\doibase 10.1103/PhysRevB.95.115425} {\bibfield  {journal}
  {\bibinfo  {journal} {Phys. Rev. B}\ }\textbf {\bibinfo {volume} {95}},\
  \bibinfo {pages} {115425} (\bibinfo {year} {2017})}\BibitemShut {NoStop}%
\bibitem [{\citenamefont {Sun}\ \emph {et~al.}()\citenamefont {Sun},
  \citenamefont {Basov},\ and\ \citenamefont {Fogler}}]{Sun2017}%
  \BibitemOpen
  \bibfield  {author} {\bibinfo {author} {\bibfnamefont {Z.}~\bibnamefont
  {Sun}}, \bibinfo {author} {\bibfnamefont {D.~N.}\ \bibnamefont {Basov}}, \
  and\ \bibinfo {author} {\bibfnamefont {M.~M.}\ \bibnamefont {Fogler}},\
  }\href {http://arxiv.org/abs/1704.07334} {\ }\bibinfo {note}
  {(unpublished)},\ \Eprint {http://arxiv.org/abs/1704.07334}
  {arXiv:1704.07334} \BibitemShut {NoStop}%
\bibitem [{\citenamefont {Guo}\ \emph {et~al.}(2017)\citenamefont {Guo},
  \citenamefont {Ilseven}, \citenamefont {Falkovich},\ and\ \citenamefont
  {Levitov}}]{Guo2017}%
  \BibitemOpen
  \bibfield  {author} {\bibinfo {author} {\bibfnamefont {H.}~\bibnamefont
  {Guo}}, \bibinfo {author} {\bibfnamefont {E.}~\bibnamefont {Ilseven}},
  \bibinfo {author} {\bibfnamefont {G.}~\bibnamefont {Falkovich}}, \ and\
  \bibinfo {author} {\bibfnamefont {L.~S.}\ \bibnamefont {Levitov}},\ }\href
  {\doibase 10.1073/pnas.1612181114} {\bibfield  {journal} {\bibinfo  {journal}
  {Proc. Natl. Acad. Sci.}\ }\textbf {\bibinfo {volume} {114}},\ \bibinfo
  {pages} {3068} (\bibinfo {year} {2017})}\BibitemShut {NoStop}%
\bibitem [{\citenamefont {Mikhailov}(2007)}]{Mikhailov2007}%
  \BibitemOpen
  \bibfield  {author} {\bibinfo {author} {\bibfnamefont {S.~A.}\ \bibnamefont
  {Mikhailov}},\ }\href {\doibase 10.1209/0295-5075/79/27002} {\bibfield
  {journal} {\bibinfo  {journal} {Europhys. Lett.}\ }\textbf {\bibinfo {volume}
  {79}},\ \bibinfo {pages} {27002} (\bibinfo {year} {2007})}\BibitemShut
  {NoStop}%
\bibitem [{\citenamefont {Mikhailov}(2011)}]{Mikhailov2011}%
  \BibitemOpen
  \bibfield  {author} {\bibinfo {author} {\bibfnamefont {S.~A.}\ \bibnamefont
  {Mikhailov}},\ }\href {\doibase 10.1103/PhysRevB.84.045432} {\bibfield
  {journal} {\bibinfo  {journal} {Phys. Rev. B}\ }\textbf {\bibinfo {volume}
  {84}},\ \bibinfo {pages} {045432} (\bibinfo {year} {2011})}\BibitemShut
  {NoStop}%
\bibitem [{\citenamefont {Gullans}\ \emph {et~al.}(2013)\citenamefont
  {Gullans}, \citenamefont {Chang}, \citenamefont {Koppens}, \citenamefont
  {de~Abajo},\ and\ \citenamefont {Lukin}}]{Gullans2013}%
  \BibitemOpen
  \bibfield  {author} {\bibinfo {author} {\bibfnamefont {M.}~\bibnamefont
  {Gullans}}, \bibinfo {author} {\bibfnamefont {D.~E.}\ \bibnamefont {Chang}},
  \bibinfo {author} {\bibfnamefont {F.~H.~L.}\ \bibnamefont {Koppens}},
  \bibinfo {author} {\bibfnamefont {F.~J.~G.}\ \bibnamefont {de~Abajo}}, \ and\
  \bibinfo {author} {\bibfnamefont {M.~D.}\ \bibnamefont {Lukin}},\ }\href
  {\doibase 10.1103/PhysRevLett.111.247401} {\bibfield  {journal} {\bibinfo
  {journal} {Phys. Rev. Lett.}\ }\textbf {\bibinfo {volume} {111}},\ \bibinfo
  {pages} {247401} (\bibinfo {year} {2013})}\BibitemShut {NoStop}%
\bibitem [{\citenamefont {Cheng}\ \emph {et~al.}(2014)\citenamefont {Cheng},
  \citenamefont {Vermeulen},\ and\ \citenamefont {Sipe}}]{Cheng2014}%
  \BibitemOpen
  \bibfield  {author} {\bibinfo {author} {\bibfnamefont {J.~L.}\ \bibnamefont
  {Cheng}}, \bibinfo {author} {\bibfnamefont {N.}~\bibnamefont {Vermeulen}}, \
  and\ \bibinfo {author} {\bibfnamefont {J.~E.}\ \bibnamefont {Sipe}},\ }\href
  {\doibase 10.1088/1367-2630/16/5/053014} {\bibfield  {journal} {\bibinfo
  {journal} {New J. Phys.}\ }\textbf {\bibinfo {volume} {16}},\ \bibinfo
  {pages} {053014} (\bibinfo {year} {2014})}\BibitemShut {NoStop}%
\bibitem [{\citenamefont {Yao}\ \emph {et~al.}(2014)\citenamefont {Yao},
  \citenamefont {Tokman},\ and\ \citenamefont {Belyanin}}]{Yao2014}%
  \BibitemOpen
  \bibfield  {author} {\bibinfo {author} {\bibfnamefont {X.}~\bibnamefont
  {Yao}}, \bibinfo {author} {\bibfnamefont {M.}~\bibnamefont {Tokman}}, \ and\
  \bibinfo {author} {\bibfnamefont {A.}~\bibnamefont {Belyanin}},\ }\href
  {\doibase 10.1103/PhysRevLett.112.055501} {\bibfield  {journal} {\bibinfo
  {journal} {Phys. Rev. Lett.}\ }\textbf {\bibinfo {volume} {112}},\ \bibinfo
  {pages} {055501} (\bibinfo {year} {2014})}\BibitemShut {NoStop}%
\bibitem [{\citenamefont {Cheng}\ \emph {et~al.}(2015)\citenamefont {Cheng},
  \citenamefont {Vermeulen},\ and\ \citenamefont {Sipe}}]{Cheng2015}%
  \BibitemOpen
  \bibfield  {author} {\bibinfo {author} {\bibfnamefont {J.~L.}\ \bibnamefont
  {Cheng}}, \bibinfo {author} {\bibfnamefont {N.}~\bibnamefont {Vermeulen}}, \
  and\ \bibinfo {author} {\bibfnamefont {J.~E.}\ \bibnamefont {Sipe}},\ }\href
  {\doibase 10.1103/PhysRevB.91.235320} {\bibfield  {journal} {\bibinfo
  {journal} {Phys. Rev. B}\ }\textbf {\bibinfo {volume} {91}},\ \bibinfo
  {pages} {235320} (\bibinfo {year} {2015})}\BibitemShut {NoStop}%
\bibitem [{\citenamefont {Mikhailov}(2016)}]{Mikhailov2016}%
  \BibitemOpen
  \bibfield  {author} {\bibinfo {author} {\bibfnamefont {S.~A.}\ \bibnamefont
  {Mikhailov}},\ }\href {\doibase 10.1103/PhysRevB.93.085403} {\bibfield
  {journal} {\bibinfo  {journal} {Phys. Rev. B}\ }\textbf {\bibinfo {volume}
  {93}},\ \bibinfo {pages} {085403} (\bibinfo {year} {2016})}\BibitemShut
  {NoStop}%
\bibitem [{\citenamefont {Tokman}\ \emph {et~al.}(2016)\citenamefont {Tokman},
  \citenamefont {Wang}, \citenamefont {Oladyshkin}, \citenamefont {Kutayiah},\
  and\ \citenamefont {Belyanin}}]{Tokman2016}%
  \BibitemOpen
  \bibfield  {author} {\bibinfo {author} {\bibfnamefont {M.}~\bibnamefont
  {Tokman}}, \bibinfo {author} {\bibfnamefont {Y.}~\bibnamefont {Wang}},
  \bibinfo {author} {\bibfnamefont {I.}~\bibnamefont {Oladyshkin}}, \bibinfo
  {author} {\bibfnamefont {A.~R.}\ \bibnamefont {Kutayiah}}, \ and\ \bibinfo
  {author} {\bibfnamefont {A.}~\bibnamefont {Belyanin}},\ }\href {\doibase
  10.1103/PhysRevB.93.235422} {\bibfield  {journal} {\bibinfo  {journal} {Phys.
  Rev. B}\ }\textbf {\bibinfo {volume} {93}},\ \bibinfo {pages} {235422}
  (\bibinfo {year} {2016})}\BibitemShut {NoStop}%
\bibitem [{\citenamefont {Cheng}\ \emph {et~al.}(2017)\citenamefont {Cheng},
  \citenamefont {Vermeulen},\ and\ \citenamefont {Sipe}}]{Cheng2016}%
  \BibitemOpen
  \bibfield  {author} {\bibinfo {author} {\bibfnamefont {J.~L.}\ \bibnamefont
  {Cheng}}, \bibinfo {author} {\bibfnamefont {N.}~\bibnamefont {Vermeulen}}, \
  and\ \bibinfo {author} {\bibfnamefont {J.~E.}\ \bibnamefont {Sipe}},\ }\href
  {\doibase 10.1038/srep43843} {\bibfield  {journal} {\bibinfo  {journal} {Sci.
  Rep.}\ }\textbf {\bibinfo {volume} {7}},\ \bibinfo {pages} {43843} (\bibinfo
  {year} {2017})}\BibitemShut {NoStop}%
\bibitem [{\citenamefont {Manzoni}\ \emph {et~al.}()\citenamefont {Manzoni},
  \citenamefont {Silveiro}, \citenamefont {Abajo},\ and\ \citenamefont
  {Chang}}]{Manzoni}%
  \BibitemOpen
  \bibfield  {author} {\bibinfo {author} {\bibfnamefont {M.~T.}\ \bibnamefont
  {Manzoni}}, \bibinfo {author} {\bibfnamefont {I.}~\bibnamefont {Silveiro}},
  \bibinfo {author} {\bibfnamefont {F.~J. G.~D.}\ \bibnamefont {Abajo}}, \ and\
  \bibinfo {author} {\bibfnamefont {D.~E.}\ \bibnamefont {Chang}},\ }\href
  {\doibase 10.1088/1367-2630/17/8/083031} {\bibfield  {journal} {\bibinfo
  {journal} {New J. Phys.}\ }\textbf {\bibinfo {volume} {17}},\ \bibinfo
  {pages} {83031}}\BibitemShut {NoStop}%
\bibitem [{\citenamefont {Wang}\ \emph {et~al.}(2016)\citenamefont {Wang},
  \citenamefont {Tokman},\ and\ \citenamefont {Belyanin}}]{Wang2016}%
  \BibitemOpen
  \bibfield  {author} {\bibinfo {author} {\bibfnamefont {Y.}~\bibnamefont
  {Wang}}, \bibinfo {author} {\bibfnamefont {M.}~\bibnamefont {Tokman}}, \ and\
  \bibinfo {author} {\bibfnamefont {A.}~\bibnamefont {Belyanin}},\ }\href
  {\doibase 10.1103/PhysRevB.94.195442} {\bibfield  {journal} {\bibinfo
  {journal} {Phys. Rev. B}\ }\textbf {\bibinfo {volume} {94}},\ \bibinfo
  {pages} {195442} (\bibinfo {year} {2016})}\BibitemShut {NoStop}%
\bibitem [{\citenamefont {Rostami}\ \emph {et~al.}(2017)\citenamefont
  {Rostami}, \citenamefont {Katsnelson},\ and\ \citenamefont
  {Polini}}]{Rostami2016}%
  \BibitemOpen
  \bibfield  {author} {\bibinfo {author} {\bibfnamefont {H.}~\bibnamefont
  {Rostami}}, \bibinfo {author} {\bibfnamefont {M.~I.}\ \bibnamefont
  {Katsnelson}}, \ and\ \bibinfo {author} {\bibfnamefont {M.}~\bibnamefont
  {Polini}},\ }\href {\doibase 10.1103/PhysRevB.95.035416} {\bibfield
  {journal} {\bibinfo  {journal} {Phys. Rev. B}\ }\textbf {\bibinfo {volume}
  {95}},\ \bibinfo {pages} {035416} (\bibinfo {year} {2017})}\BibitemShut
  {NoStop}%
\bibitem [{\citenamefont {Ni}\ \emph {et~al.}(2016)\citenamefont {Ni},
  \citenamefont {Wang}, \citenamefont {Goldflam}, \citenamefont {Wagner},
  \citenamefont {Fei}, \citenamefont {McLeod}, \citenamefont {Liu},
  \citenamefont {Keilmann}, \citenamefont {{\"{O}}zyilmaz}, \citenamefont
  {{Castro Neto}}, \citenamefont {Hone}, \citenamefont {Fogler},\ and\
  \citenamefont {Basov}}]{Ni2016}%
  \BibitemOpen
  \bibfield  {author} {\bibinfo {author} {\bibfnamefont {G.~X.}\ \bibnamefont
  {Ni}}, \bibinfo {author} {\bibfnamefont {L.}~\bibnamefont {Wang}}, \bibinfo
  {author} {\bibfnamefont {M.~D.}\ \bibnamefont {Goldflam}}, \bibinfo {author}
  {\bibfnamefont {M.}~\bibnamefont {Wagner}}, \bibinfo {author} {\bibfnamefont
  {Z.}~\bibnamefont {Fei}}, \bibinfo {author} {\bibfnamefont {A.~S.}\
  \bibnamefont {McLeod}}, \bibinfo {author} {\bibfnamefont {M.~K.}\
  \bibnamefont {Liu}}, \bibinfo {author} {\bibfnamefont {F.}~\bibnamefont
  {Keilmann}}, \bibinfo {author} {\bibfnamefont {B.}~\bibnamefont
  {{\"{O}}zyilmaz}}, \bibinfo {author} {\bibfnamefont {A.~H.}\ \bibnamefont
  {{Castro Neto}}}, \bibinfo {author} {\bibfnamefont {J.}~\bibnamefont {Hone}},
  \bibinfo {author} {\bibfnamefont {M.~M.}\ \bibnamefont {Fogler}}, \ and\
  \bibinfo {author} {\bibfnamefont {D.~N.}\ \bibnamefont {Basov}},\ }\href
  {\doibase 10.1038/nphoton.2016.45} {\bibfield  {journal} {\bibinfo  {journal}
  {Nature Photon.}\ }\textbf {\bibinfo {volume} {10}},\ \bibinfo {pages} {244}
  (\bibinfo {year} {2016})}\BibitemShut {NoStop}%
\bibitem [{\citenamefont {Son}\ \emph {et~al.}(2018)\citenamefont {Son},
  \citenamefont {{\v{S}}i{\v{s}}kins}, \citenamefont {Mullan}, \citenamefont
  {Yin}, \citenamefont {Kravets}, \citenamefont {Kozikov}, \citenamefont
  {Ozdemir}, \citenamefont {Alhazmi}, \citenamefont {Holwill}, \citenamefont
  {Watanabe}, \citenamefont {Taniguchi}, \citenamefont {Ghazaryan},
  \citenamefont {Novoselov}, \citenamefont {Fal'ko},\ and\ \citenamefont
  {Mishchenko}}]{Son2017}%
  \BibitemOpen
  \bibfield  {author} {\bibinfo {author} {\bibfnamefont {S.-k.}\ \bibnamefont
  {Son}}, \bibinfo {author} {\bibfnamefont {M.}~\bibnamefont
  {{\v{S}}i{\v{s}}kins}}, \bibinfo {author} {\bibfnamefont {C.}~\bibnamefont
  {Mullan}}, \bibinfo {author} {\bibfnamefont {J.}~\bibnamefont {Yin}},
  \bibinfo {author} {\bibfnamefont {V.~G.}\ \bibnamefont {Kravets}}, \bibinfo
  {author} {\bibfnamefont {A.}~\bibnamefont {Kozikov}}, \bibinfo {author}
  {\bibfnamefont {S.}~\bibnamefont {Ozdemir}}, \bibinfo {author} {\bibfnamefont
  {M.}~\bibnamefont {Alhazmi}}, \bibinfo {author} {\bibfnamefont
  {M.}~\bibnamefont {Holwill}}, \bibinfo {author} {\bibfnamefont
  {K.}~\bibnamefont {Watanabe}}, \bibinfo {author} {\bibfnamefont
  {T.}~\bibnamefont {Taniguchi}}, \bibinfo {author} {\bibfnamefont
  {D.}~\bibnamefont {Ghazaryan}}, \bibinfo {author} {\bibfnamefont {K.~S.}\
  \bibnamefont {Novoselov}}, \bibinfo {author} {\bibfnamefont {V.~I.}\
  \bibnamefont {Fal'ko}}, \ and\ \bibinfo {author} {\bibfnamefont
  {A.}~\bibnamefont {Mishchenko}},\ }\href
  {http://stacks.iop.org/2053-1583/5/i=1/a=011006} {\bibfield  {journal}
  {\bibinfo  {journal} {2D Materials}\ }\textbf {\bibinfo {volume} {5}},\
  \bibinfo {pages} {011006} (\bibinfo {year} {2018})}\BibitemShut {NoStop}%
\bibitem [{\citenamefont {Giorgianni}\ \emph {et~al.}(2016)\citenamefont
  {Giorgianni}, \citenamefont {Chiadroni}, \citenamefont {Rovere},
  \citenamefont {Cestelli-Guidi}, \citenamefont {Perucchi}, \citenamefont
  {Bellaveglia}, \citenamefont {Castellano}, \citenamefont {{Di Giovenale}},
  \citenamefont {{Di Pirro}}, \citenamefont {Ferrario}, \citenamefont
  {Pompili}, \citenamefont {Vaccarezza}, \citenamefont {Villa}, \citenamefont
  {Cianchi}, \citenamefont {Mostacci}, \citenamefont {Petrarca}, \citenamefont
  {Brahlek}, \citenamefont {Koirala}, \citenamefont {Oh},\ and\ \citenamefont
  {Lupi}}]{Lupi.2016}%
  \BibitemOpen
  \bibfield  {author} {\bibinfo {author} {\bibfnamefont {F.}~\bibnamefont
  {Giorgianni}}, \bibinfo {author} {\bibfnamefont {E.}~\bibnamefont
  {Chiadroni}}, \bibinfo {author} {\bibfnamefont {A.}~\bibnamefont {Rovere}},
  \bibinfo {author} {\bibfnamefont {M.}~\bibnamefont {Cestelli-Guidi}},
  \bibinfo {author} {\bibfnamefont {A.}~\bibnamefont {Perucchi}}, \bibinfo
  {author} {\bibfnamefont {M.}~\bibnamefont {Bellaveglia}}, \bibinfo {author}
  {\bibfnamefont {M.}~\bibnamefont {Castellano}}, \bibinfo {author}
  {\bibfnamefont {D.}~\bibnamefont {{Di Giovenale}}}, \bibinfo {author}
  {\bibfnamefont {G.}~\bibnamefont {{Di Pirro}}}, \bibinfo {author}
  {\bibfnamefont {M.}~\bibnamefont {Ferrario}}, \bibinfo {author}
  {\bibfnamefont {R.}~\bibnamefont {Pompili}}, \bibinfo {author} {\bibfnamefont
  {C.}~\bibnamefont {Vaccarezza}}, \bibinfo {author} {\bibfnamefont
  {F.}~\bibnamefont {Villa}}, \bibinfo {author} {\bibfnamefont
  {A.}~\bibnamefont {Cianchi}}, \bibinfo {author} {\bibfnamefont
  {A.}~\bibnamefont {Mostacci}}, \bibinfo {author} {\bibfnamefont
  {M.}~\bibnamefont {Petrarca}}, \bibinfo {author} {\bibfnamefont
  {M.}~\bibnamefont {Brahlek}}, \bibinfo {author} {\bibfnamefont
  {N.}~\bibnamefont {Koirala}}, \bibinfo {author} {\bibfnamefont
  {S.}~\bibnamefont {Oh}}, \ and\ \bibinfo {author} {\bibfnamefont
  {S.}~\bibnamefont {Lupi}},\ }\href {\doibase 10.1038/ncomms11421} {\bibfield
  {journal} {\bibinfo  {journal} {Nat. Commun.}\ }\textbf {\bibinfo {volume}
  {7}},\ \bibinfo {pages} {11421} (\bibinfo {year} {2016})}\BibitemShut
  {NoStop}%
\bibitem [{\citenamefont {Landau}\ and\ \citenamefont
  {Lifshitz}(1984)}]{Landau.8}%
  \BibitemOpen
  \bibfield  {author} {\bibinfo {author} {\bibfnamefont {L.~D.}\ \bibnamefont
  {Landau}}\ and\ \bibinfo {author} {\bibfnamefont {E.~M.}\ \bibnamefont
  {Lifshitz}},\ }\href@noop {} {\emph {\bibinfo {title} {{Electrodynamics of
  Continuous Media}}}}\ (\bibinfo  {publisher} {Pergamon},\ \bibinfo {address}
  {Oxford},\ \bibinfo {year} {1984})\BibitemShut {NoStop}%
\bibitem [{\citenamefont {Boyd}(2008)}]{Boyd.2008}%
  \BibitemOpen
  \bibfield  {author} {\bibinfo {author} {\bibfnamefont {R.~W.}\ \bibnamefont
  {Boyd}},\ }\href@noop {} {\emph {\bibinfo {title} {{Nonlinear optics}}}},\
  \bibinfo {edition} {3rd}\ ed.\ (\bibinfo  {publisher} {Academic Press},\
  \bibinfo {year} {2008})\BibitemShut {NoStop}%
\bibitem [{\citenamefont {Mikhailov}(2017)}]{Mikhailov2017}%
  \BibitemOpen
  \bibfield  {author} {\bibinfo {author} {\bibfnamefont {S.~A.}\ \bibnamefont
  {Mikhailov}},\ }\href {\doibase 10.1021/acsphotonics.7b00468} {\bibfield
  {journal} {\bibinfo  {journal} {ACS Photonics}\ } (\bibinfo {year} {2017}),\
  10.1021/acsphotonics.7b00468}\BibitemShut {NoStop}%
\bibitem [{\citenamefont {Nielsen}\ and\ \citenamefont
  {Shklovskii}(1969)}]{SHKLOSVKII.1969}%
  \BibitemOpen
  \bibfield  {author} {\bibinfo {author} {\bibfnamefont {H.}~\bibnamefont
  {Nielsen}}\ and\ \bibinfo {author} {\bibfnamefont {B.~I.}\ \bibnamefont
  {Shklovskii}},\ }\href
  {http://www.jetp.ac.ru/cgi-bin/e/index/r/56/2/p709?a=list} {\bibfield
  {journal} {\bibinfo  {journal} {Sov. Phys. JETP}\ }\textbf {\bibinfo {volume}
  {56}},\ \bibinfo {pages} {709} (\bibinfo {year} {1969})}\BibitemShut
  {NoStop}%
\end{thebibliography}%

\end{document}